\documentclass[thmsa,letterpaper,12pt]{article}
\usepackage{amsfonts}
\usepackage{amsmath}
\usepackage{sw20lart}

\input tcilatex
\begin{document}

\author{Nick Laskin\thanks{%
E-mail address: nlaskin@rocketmail.com}}
\title{\textbf{New Pricing Framework: }\\
\textbf{Options and Bonds }\\
}
\date{TopQuark Inc.\\
Toronto, ON, M6P 2P2}
\maketitle

\begin{abstract}
A unified analytical pricing framework with involvement of the shot noise
random process has been introduced and elaborated. Two exactly solvable new
models have been developed.

The first model has been designed to value options. It is assumed that asset
price stochastic dynamics follows a Geometric Shot Noise motion. A new
arbitrage-free integro-differential option pricing equation has been found
and solved. The put-call parity has been proved and the Greeks have been
calculated. Three additional new Greeks associated with market model
parameters have been introduced and evaluated. It has been shown that in
diffusion approximation the developed option pricing model incorporates the
well-known Black-Scholes equation and its solution. The stochastic dynamic
origin of the Black-Scholes volatility has been uncovered.

The new option pricing model has been generalized based on asset price
dynamics modeled by the superposition of Geometric Brownian motion and
Geometric Shot Noise. A generalized arbitrage-free integro-differential
option pricing equation has been obtained and solved. Based on this solution
new generalized Greeks have been introduced and calculated.

To model stochastic dynamics of a short term interest rate, the second model
has been introduced and developed based on Langevin type equation with shot
noise. It has been found that the model provides affine term structure. A
new bond pricing formula has been obtained. It has been shown that in
diffusion approximation the developed bond pricing formula goes into the
well-known Vasi\v{c}ek solution. The stochastic dynamic origin of the
long-term mean and instantaneous volatility of the Vasi\v{c}ek model has
been uncovered.

A generalized bond pricing model has been introduced and developed based on
short term interest rate stochastic dynamics modeled by superposition of a
standard Wiener process and shot noise.

Despite the non-Gaussianity of probability distributions involved, all newly
elaborated models have the same degree of analytical tractability as the
Black--Scholes model and the Vasi\v{c}ek model. This circumstance allows one
to derive simple exact formulae to value options and bonds.

\textit{Key words}: Financial derivatives fundamentals, Shot noise, Option
pricing equation, Green function, Put-call parity, Greeks, Black-Scholes
equation, Short term interest rate, Vasi\v{c}ek model, Affine term
structure, Bond pricing formula.
\end{abstract}

\tableofcontents

\section{Introduction}

The aim of this paper is to introduce and elaborate a new unified analytical
framework to value options and bonds.

The options pricing approach is based on asset price dynamics that has been
modelled by the stochastic differential equation with involvement of shot
noise. It results in Geometric Shot Noise motion of asset price. A new
arbitrage-free integro-differential option pricing equation has been
developed and solved. New exact formulas to value European call and put
options have been obtained. The put-call parity has been proved. The Greeks
have been calculated based on the solution of the option pricing equation.
Three new Greeks associated with the market model parameters have been
introduced and evaluated. It has been shown that the developed option
pricing framework incorporates the well-known Black-Scholes equation \cite%
{Black-Scholes}. The Black-Scholes equation and its solutions emerge from
our integro-differential option pricing equation in the special case which
we call "diffusion approximation". The Geometric Shot Noise model in
diffusion approximation explains the stochastic dynamic origin of volatility
in the Black-Scholes model.

The bonds pricing analytical approach is based on the Langevin type
stochastic differential equation with shot noise to model a short term
interest rate dynamics. It results in non-Gaussian random motion of short
term interest rate. A bond pricing formula has been obtained and it has been
shown that the model provides affine term structure. The new bond pricing
formula incorporates the well-known Vasi\v{c}ek solution \cite{Vasicek}. The
Vasi\v{c}ek solution comes out from our bond pricing formula in diffusion
approximation. The stochastic dynamic origin of the Vasi\v{c}ek long-term
mean and instantaneous volatility has been uncovered.

The paper's main results are presented by Eqs.(\ref{eq1}), (\ref{eq2}), (\ref%
{eq16}), (\ref{eq22}), (\ref{eq39}), (\ref{eq43}), (\ref{eq113})-(\ref{eq116}%
), (\ref{eq124}), (\ref{eq131}), (\ref{eq136}), (\ref{eq142}), (\ref{eq144}%
), (\ref{eq162}), (\ref{eq166}) and (\ref{eq175}).

New formulae to evaluate the common Greeks for call and put options have
been obtained based on an exact solution of the integro-differential option
pricing equation. The formulae are listed in Table 1 and Table 2.

The paper is organized as follows.

In Sec.2 Geometric Shot Noise motion has been introduced and applied to
model asset price dynamics.

A new arbitrage-free integro-differential equation to value options is
obtained in Sec.3. It has been shown that Green's function method is an
effective mathematical tool to solve this equation. The exact analytical
solutions to this equation have been found for European call and put options.

The put-call parity has been proved in Sec.4.

The Greeks, including three newly introduced Greeks, have been calculated in
Sec.5. Three new Greeks are sensitivities associated with the market
parameters involved into the definition of the shot noise process. The
Gaussian model for asset price jumps has been considered to find the
formulae for three new Greeks.

The diffusion approximation of the integro-differential option pricing
equation has been defined and elaborated in Sec.6. It has been shown that
the well-known Black-Scholes equation and its volatility come out from the
integro-differential pricing equation in diffusion approximation. It has
been shown as well that the solution to the Black-Scholes equation
straightforwardly follows from the exact solution to the
integro-differential pricing equation. The well-known Black-Scholes Greeks
for European call options have been replicated from the new common Greeks in
diffusion approximation. The Black-Scholes Greeks for European call options
are summarized in Table 3.

The generalized option pricing framework has been presented in Sec.7. The
generalization comes from the idea to accommodate a superposition of
Geometric Brownian motion and Geometric Shot Noise in the equation for asset
price dynamics. The outcome of implementing this idea is a generalized
pricing equation. New formulae to value European call and put options have
been obtained as solutions to the generalized pricing equation. The special
limit cases of those formulas have been developed and discussed. The
generalized Greeks have been introduced and calculated. Table 4 displays new
formulas for the generalized Greeks.

In Sec.8 short term interest rate has been modeled by the Langevin
stochastic differential equation with shot noise. A new bond pricing formula
has been found, and it has been shown that the new model provides affine
term structure. The bond pricing formula is the solution to the
integro-differential term structure equation. The well-known Vasi\v{c}ek
model for short term interest rate with its long-term mean and instantaneous
volatility comes out from our model in diffusion approximation. It has been
shown as well that the Vasi\v{c}ek bond pricing formula comes out from the
new bond pricing formula in diffusion approximation. It has to be emphasized
that our model is non-Gaussian, while the Vasi\v{c}ek model is Gaussian,
that is, probability distributions of short term interest rate are different
for our model and the Vasi\v{c}ek model. It is interesting, that despite
this difference both models possess exactly the same mean and variance.

A generalized bond pricing model has been introduced and developed in Sec.9
based on short term interest rate stochastic dynamics modeled by
superposition of a standard Wiener process and shot noise. A new bond
pricing formula has been found and it has been shown that the generalized
model provides affine term structure. The generalized term structure
equation has been found and solved.

It has to be emphasized that despite the non-Gaussianity of probability
distributions involved, all newly developed quantitative models to value
options and bonds have the same degree of analytical tractability as the
Black--Scholes model \cite{Black-Scholes} and the Vasi\v{c}ek model \cite%
{Vasicek}. Analytical tractability allows one to obtain new exact simple
formulas to value options and bonds.

The paper's main results and findings have been summarized and discussed in
the Conclusion.

Appendix A develops Green's function method to solve the Black-Scholes
equation for European call options. The well-known solution has been
obtained straightforwardly without preliminary transformations to convert
the Black-Scholes equation to the heat equation. To our best knowledge, the
implemented Green's function approach has not yet been presented anywhere.

Appendix B is a walkthrough to show that in diffusion approximation the
equation for the new Greek $\Theta _{C}$ goes into the well-known formula
for the Black-Scholes Greek theta.

\section{Asset price stochastic dynamics}

\subsection{Geometric Shot Noise motion}

It is supposed that asset price $S(t)$ follows the stochastic differential
equation

\begin{equation}
dS=SF(t)dt,  \label{eq1}
\end{equation}

with random force $F(t)$ modeled by the shot noise process (see, Eq.(6) in 
\cite{Laskin})

\begin{equation}
F(t)=\dsum\limits_{k=1}^{n}\eta _{k}\varphi (t-t_{k}),  \label{eq2}
\end{equation}

where random jumps $\eta _{k}$ of asset price are statistically independent
and distributed with probability density function $p(\eta )$, random time
points $t_{k}$, which are arrival times of asset price jumps, are uniformly
distributed on time interval $[t,T]$, so that their total number $n$ obeys
the Poisson law with parameter $\lambda $, and deterministic function $%
\varphi (t)$ is the response function.

It is supposed that defined by Eq.(\ref{eq2}) shot noise process $F(t)$
describes the influence of different fluctuating factors on asset price
dynamics. A single shot noise pulse $\eta _{k}\varphi (t-t_{k})$ describes
the influence of a piece of information which has become available at random
moment $t_{k}$ on the asset price at a later time $t$. Amplitude $\eta _{k}$
responds to the magnitude of the asset price pulse $\varphi (t-t_{k})$.
Amplitudes $\eta _{k}$ are random statistically independent variables,
subject to the market information available. We assume as well that each
pulse has the same functional form or, in other words, one general response
function $\varphi (t)$ can be used to describe the asset price dynamics.

We call the stochastic dynamics introduced by Eqs.(\ref{eq1}) and (\ref{eq2}%
) Geometric Shot Noise motion.

\subsection{Characteristic functional of shot noise}

By definition the characteristic functional $\Phi \lbrack \alpha (\tau )]$
of random force $F(t)$ is%
\begin{equation}
\Phi \lbrack \alpha (\tau )]=\left\langle \exp \left\{
i\dint\limits_{t}^{T}d\tau \alpha (\tau )F(\tau )\right\} \right\rangle ,
\label{eq3}
\end{equation}

where $\alpha (\tau )$ is an arbitrary sufficiently smooth function and $%
\left\langle ...\right\rangle $ stands for the average over all randomness
involved into the random force $F(t)$.

The characteristic functional contains all information about statistical
moments of random force $F(t)$. For example, the mean value of $F(t)$ can be
calculated as functional derivative

\begin{equation}
<F(t)>=\frac{1}{i}\frac{\delta }{\delta \alpha (t)}\Phi \lbrack \alpha (\tau
)]|_{\alpha (\tau )=0},  \label{eq4}
\end{equation}

while the correlation function $<F(t_{1})F(t_{2})>$ is given by the second
order functional derivative

\begin{equation}
<F(t_{1})F(t_{2})>=\frac{1}{i^{2}}\frac{\delta ^{2}}{\delta \alpha
(t_{1})\delta \alpha (t_{2})}\Phi \lbrack \alpha (\tau )]|_{\alpha (\tau
)=0}.  \label{eq5}
\end{equation}

To evaluate the characteristic functional $\Phi \lbrack \alpha (\tau )]$ we
need to define probabilistic characteristics of each of three sources of
randomness involved in Eq.(\ref{eq2}). Therefore, assuming that these three
sources of randomness are independent of each other, we have three
statistically independent averaging procedures \cite{Laskin}:

1. Averaging over uniformly distributed time points $t_{k}$ on the interval $%
[t,T]$,

\begin{equation}
\left\langle ...\right\rangle _{T-t}=\dprod\limits_{k=1}^{n}\frac{1}{T-t}%
\dint\limits_{t}^{T}dt_{k}....  \label{eq6}
\end{equation}

2. Averaging over random asset price jumps, which are statistically
independent and distributed with probability density function $p(\eta )$,

\begin{equation}
\left\langle ...\right\rangle _{\eta
}=\dprod\limits_{k=1}^{n}\dint\limits_{-\infty }^{\infty }d\eta _{k}p(\eta
_{k})....  \label{eq7}
\end{equation}

3. Averaging over random number $n$ of price jumps,

\begin{equation}
\left\langle ...\right\rangle _{n}=e^{-\lambda
(T-t)}\dsum\limits_{n=0}^{\infty }\frac{(\lambda (T-t))^{n}}{n!}...,
\label{eq8}
\end{equation}

which is in fact the averaging with the Poisson probability density
function, and $\lambda $ is the rate of arrival of price jumps, i.e. the
number of jumps per unit time.

Now we are in position to calculate $\Phi \lbrack \alpha (\tau )]$. First,
by performing steps \#1 and \#2 we obtain

\begin{equation}
\Phi \lbrack \alpha (\tau )]=\left\langle \exp \left\{
i\dint\limits_{t}^{T}d\tau \alpha (\tau )\dsum\limits_{k=1}^{n}\eta
_{k}\varphi (\tau -t_{k})\right\} \right\rangle _{t_{k},\eta }  \label{eq9}
\end{equation}

\begin{equation*}
=\left[ \frac{1}{T-t}\dint\limits_{t}^{T}dt^{\prime }\dint\limits_{-\infty
}^{\infty }d\eta p(\eta )\exp \left\{ i\eta \dint\limits_{t^{\prime
}}^{T}d\tau \alpha (\tau )\varphi (\tau -t^{\prime })\right\} \right] ^{n}.
\end{equation*}

Then, let us do step \#3,

\begin{equation*}
\Phi \lbrack \alpha (\tau )]=\left\langle \left[ \frac{1}{T-t}%
\dint\limits_{t}^{T}dt^{\prime }\dint\limits_{-\infty }^{\infty }d\eta
p(\eta )\exp \left\{ i\eta \dint\limits_{t^{\prime }}^{T}d\tau \alpha (\tau
)\varphi (\tau -t^{\prime })\right\} \right] ^{n}\right\rangle _{n}
\end{equation*}

\begin{equation}
=e^{-\lambda (T-t)}\dsum\limits_{n=0}^{\infty }\frac{(\lambda (T-t))^{n}}{n!}%
{\LARGE [}\frac{1}{T-t}\dint\limits_{t}^{T}dt^{\prime }  \label{eq10}
\end{equation}

\begin{equation*}
\times \dint\limits_{-\infty }^{\infty }d\eta p(\eta )\exp \left\{ i\eta
\dint\limits_{t^{\prime }}^{T}d\tau \alpha (\tau )\varphi (\tau -t^{\prime
})\right\} {\LARGE ]}^{n}
\end{equation*}

\begin{equation*}
=\exp \left\{ \lambda \dint\limits_{t}^{T}dt^{\prime }\dint\limits_{-\infty
}^{\infty }d\eta p(\eta )\exp \{i\eta \dint\limits_{t^{\prime }}^{T}d\tau
\alpha (\tau )\varphi (\tau -t^{\prime })\}-1\right\} .
\end{equation*}

Hence, we found the equation for the characteristic functional $\Phi \lbrack
\alpha (\tau )]$ defined by Eq.(\ref{eq3})

\begin{equation}
\Phi \lbrack \alpha (\tau )]=\exp \left\{ \lambda
\dint\limits_{t}^{T}dt^{\prime }\dint\limits_{-\infty }^{\infty }d\eta
p(\eta )\exp \{i\eta \dint\limits_{t^{\prime }}^{T}d\tau \alpha (\tau
)\varphi (\tau -t^{\prime })\}-1\right\} .  \label{eq11}
\end{equation}

Further, we chose, as an example, the response function

\begin{equation}
\varphi (t)=\delta (t),  \label{eq12}
\end{equation}

where $\delta (t)$ is delta-function. In this case we have for the
characteristic functional $\Phi \lbrack \alpha (\tau )]$

\begin{equation}
\Phi \lbrack \alpha (\tau )]=\exp \left\{ \lambda \dint\limits_{t}^{T}d\tau
\dint\limits_{-\infty }^{\infty }d\eta p(\eta )(e^{i\eta \alpha (\tau
)}-1)\right\} .  \label{eq13}
\end{equation}

Now, having Eq.(\ref{eq13}) and definitions (\ref{eq4}) and (\ref{eq5}) one
can easily obtain the mean $<F(t)>$

\begin{equation}
<F(t)>=\lambda \dint\limits_{-\infty }^{\infty }d\eta p(\eta )\eta ,
\label{eq14}
\end{equation}

and the correlation function $<F(t_{1})F(t_{2})>$

\begin{equation}
<F(t_{1})F(t_{2})>=\lambda \left( \dint\limits_{-\infty }^{\infty }d\eta
p(\eta )\eta ^{2}\right) \delta (t_{1}-t_{2})+<F(t_{1})><F(t_{2})>.
\label{eq15}
\end{equation}

\section{Option pricing equation and its solutions}

\subsection{A new arbitrage-free integro-differential pricing equation}

Having the characteristic functional $\Phi \lbrack \alpha (\tau )]$ given by
Eq.(\ref{eq13}) and assuming a frictionless and no-arbitrage market, a
constant risk-free interest rate $r$, and asset price dynamics governed by
the Geometric Shot Noise motion given by Eq.(\ref{eq1}), we introduce a new
arbitrage-free integro-differential option pricing equation

\begin{equation}
\frac{\partial C(x,t)}{\partial t}+(r-q)\frac{\partial C(x,t)}{\partial x}
\label{eq16}
\end{equation}

\begin{equation*}
+\lambda \dint\limits_{-\infty }^{\infty }d\eta p(\eta )\left\{ C(x+\eta
,t)-C(x,t)-(e^{\eta }-1)\frac{\partial C(x,t)}{\partial x}\right\}
-rC(x,t)=0,
\end{equation*}

where

\begin{equation}
x=\ln \frac{S}{K},  \label{eq17}
\end{equation}

and $C(x,t)$ is the value of a European call option\footnote{%
An option which gives the owner the right, but not the obligation, to buy an
asset, at a specified price (strike price $K$), by a predetermined date
(maturity time $T$).} on divident-paying asset, $S$ is asset price governed
by Eq.(\ref{eq1}), $K$ is the strike price, $r$ is the risk-free interest
rate, $q$ is continuously paid dividend yield, which is a constant, $p(\eta
) $ is the probability density function involved into Eq.(\ref{eq2}).

The terminal condition (or payoff function) for a European call option is

\begin{equation}
C(S,T)=\max (S-K,0),  \label{eq18}
\end{equation}

where $T$ is an option maturity time.

If we take into account Eq.(\ref{eq17}), then we can write Eq.(\ref{eq18}) as

\begin{equation}
C(x,T)=K\max (e^{x}-1,0).  \label{eq19}
\end{equation}

Thus, the new generalized option pricing framework has been introduced by
Eqs.(\ref{eq16}) and (\ref{eq19}).

To value a European put option\footnote{%
An option which gives the owner the right, but not the obligation, to sell
an asset, at a strike price $K$, on the maturity date $T$.} $P(x,t)$ we have
the same equation as Eq.(\ref{eq16}) while the terminal condition is

\begin{equation}
P(S,T)=\max (K-S,0).  \label{eq20}
\end{equation}

With help of Eq.(\ref{eq17}) the terminal condition (\ref{eq20}) for a
European put option becomes

\begin{equation}
P(x,T)=K\max (1-e^{x},0).  \label{eq21}
\end{equation}

If we go from $C(x,t)$ to $C(S,t)$, where $x$ and $S$ are related to each
other through Eq.(\ref{eq17}), then we can write an option pricing equation (%
\ref{eq16}) in the form

\begin{equation}
\frac{\partial C(S,t)}{\partial t}+(r-q)S\frac{\partial C(S,t)}{\partial S}
\label{eq22}
\end{equation}

\begin{equation*}
+\lambda \dint\limits_{-\infty }^{\infty }d\eta p(\eta )\left\{ C(Se^{\eta
},t)-C(S,t)-(e^{\eta }-1)S\frac{\partial C(S,t)}{\partial S}\right\}
-rC(S,t)=0,
\end{equation*}

with the terminal condition given either by Eq.(\ref{eq18}) or by Eq.(\ref%
{eq20}).

When valuing an option, it is common practice to consider time to expiry $%
T-t $ instead of time $t$. Taking into account that

\begin{equation}
\frac{\partial }{\partial t}\rightarrow -\frac{\partial }{\partial (T-t)},
\label{eq22T}
\end{equation}

we can present Eq.(\ref{eq16}) in the form

\begin{equation*}
-\frac{\partial C(x,T-t)}{\partial (T-t)}+(r-q)\frac{\partial C(x,T-t)}{%
\partial x}
\end{equation*}

\begin{equation}
+\lambda \dint\limits_{-\infty }^{\infty }d\eta p(\eta )\left\{ C(x+\eta
,T-t)-C(x,T-t)-(e^{\eta }-1)\frac{\partial C(x,T-t)}{\partial x}\right\}
\label{eq23T}
\end{equation}

\begin{equation*}
-rC(x,T-t)=0,
\end{equation*}

while the terminal condition (\ref{eq19}) for a call option becomes

\begin{equation}
C(x,T-t)|_{t=T}=C(x,0)=K\max (e^{x}-1,0),  \label{eq24T}
\end{equation}

and the terminal condition (\ref{eq21}) for a put option becomes

\begin{equation}
P(x,T-t)|_{t=T}=P(x,0)=K\max (1-e^{x},0).  \label{eq25T}
\end{equation}

\subsection{Exact solution to the integro-differential equation}

\subsubsection{Call option}

To solve Eq.(\ref{eq23T}) subject to terminal condition given by (\ref{eq24T}%
) we will use the Green's function method.

By definition, Green's function $G(x-x^{\prime },T-t)$ satisfies the
integro-differential equation

\begin{equation*}
-\frac{\partial G(x-x^{\prime },T-t)}{\partial (T-t)}+(r-q)\frac{\partial
G(x-x^{\prime },T-t)}{\partial x}
\end{equation*}

\begin{equation}
+\lambda \dint\limits_{-\infty }^{\infty }d\eta p(\eta ){\LARGE \{}G(x+\eta
-x^{\prime },T-t)-G(x-x^{\prime },T-t)  \label{eq23}
\end{equation}

\begin{equation*}
-(e^{\eta }-1)\frac{\partial G(x-x^{\prime },T-t)}{\partial x}{\LARGE \}}%
-rG(x-x^{\prime },T-t)=0,
\end{equation*}

and the terminal condition

\begin{equation}
G(x-x^{\prime },T-t)|_{t=T}=G(x-x^{\prime },0)=\delta (x-x^{\prime }),
\label{eq24}
\end{equation}

where $T-t$ is the time to maturity.

Having Green's function $G(x-x^{\prime },T-t),$ we write the solution to Eq.(%
\ref{eq23T}) with condition (\ref{eq24T}) in the form

\begin{equation}
C(x,T-t)=K\dint\limits_{-\infty }^{\infty }dx^{\prime }G(x-x^{\prime
},T-t)\max (e^{x^{\prime }}-1,0),  \label{eq25}
\end{equation}

and the solution to Eq.(\ref{eq23T}) with condition (\ref{eq25T}) in the form

\begin{equation}
P(x,T-t)=K\dint\limits_{-\infty }^{\infty }dx^{\prime }G(x-x^{\prime
},T-t)\max (1-e^{x^{\prime }},0).  \label{eq26}
\end{equation}

Green's function introduced by Eqs.(\ref{eq23}) and (\ref{eq24}) can be
found by the Fourier transform method. Green's function $G(x-x^{\prime
},T-t) $ reads

\begin{equation}
G(x-x^{\prime },T-t)=\frac{1}{2\pi }\dint\limits_{-\infty }^{\infty
}dke^{ik(x-x^{\prime })}G(k,T-t),  \label{eq27}
\end{equation}

where $G(k,T-t)$ is the Fourier transform of Green's function defined by

\begin{equation}
G(k,T-t)=\dint\limits_{-\infty }^{\infty }dxe^{-ikx}G(x,T-t).  \label{eq28}
\end{equation}

In terms of $G(k,T-t)$, Eq.(\ref{eq23}) takes the form

\begin{equation}
\frac{\partial G(k,T-t)}{\partial (T-t)}=[-r+ik(r-q)  \label{eq29}
\end{equation}

\begin{equation*}
+\lambda \dint\limits_{-\infty }^{\infty }d\eta p(\eta )\left\{ e^{ik\eta
}-1-ik(e^{\eta }-1)\right\} ]G(k,T-t),
\end{equation*}

while the terminal condition (\ref{eq24}) becomes

\begin{equation}
G(k,T-t)|_{t=T}=G(k,0)=1.  \label{eq30}
\end{equation}

The solution to the problem defined by Eqs.(\ref{eq29}) and (\ref{eq30}) is

\begin{equation}
G(k,T-t)=e^{-r(T-t)}  \label{eq31}
\end{equation}

\begin{equation*}
\times \exp \left\{ [ik(r-q)+\lambda \dint\limits_{-\infty }^{\infty }d\eta
p(\eta )\left\{ e^{ik\eta }-1-ik(e^{\eta }-1)\right\} ](T-t)\right\} ,
\end{equation*}

Then Eq.(\ref{eq27}) gives us Green's function $G(x-x^{\prime },T-t)$

\begin{equation}
G(x-x^{\prime },T-t)=\frac{e^{-r(T-t)}}{2\pi }\dint\limits_{-\infty
}^{\infty }dke^{ik(x-x^{\prime })}  \label{eq32}
\end{equation}

\begin{equation*}
\times \exp \left\{ [ik(r-q)+\lambda \dint\limits_{-\infty }^{\infty }d\eta
p(\eta )\left\{ e^{ik\eta }-1-ik(e^{\eta }-1)\right\} ](T-t)\right\} .
\end{equation*}

Substituting Eq.(\ref{eq32}) into Eq.(\ref{eq25}) yields for the value of a
European call option $C(x,T-t)$

\begin{equation*}
C(x,T-t)=\frac{Ke^{-r(T-t)}}{2\pi }\dint\limits_{-\infty }^{\infty
}dx^{\prime }\dint\limits_{-\infty }^{\infty }dke^{ik(x-x^{\prime })}
\end{equation*}

\begin{equation}
\times \exp \left\{ [ik(r-q)+\lambda \dint\limits_{-\infty }^{\infty }d\eta
p(\eta )\left\{ e^{ik\eta }-1-ik(e^{\eta }-1)\right\} ](T-t)\right\}
\label{eq33}
\end{equation}

\begin{equation*}
\times \max (e^{x^{\prime }}-1,0).
\end{equation*}

Further, changing integration variable $x^{\prime }$ to $z$

\begin{equation}
z=x-x^{\prime }+(r-q-\lambda \varsigma )(T-t),  \label{eq34}
\end{equation}

and introducing parameter $l$ defined by

\begin{equation}
l=x+(r-q-\lambda \varsigma )(T-t),  \label{eq35}
\end{equation}

yield

\begin{equation}
C(x,T-t)=Ke^{-r(T-t)}\frac{e^{l}}{2\pi }\dint\limits_{-\infty
}^{l}dz\dint\limits_{-\infty }^{\infty }dke^{ikz-z}\exp \left\{ \lambda
(T-t)\xi (k)\right\}  \label{eq36}
\end{equation}

\begin{equation*}
-\frac{Ke^{-r(T-t)}}{2\pi }\dint\limits_{-\infty
}^{l}dz\dint\limits_{-\infty }^{\infty }dke^{ikz}\exp \left\{ \lambda
(T-t)\xi (k)\right\} ,
\end{equation*}

where $\varsigma $ and $\xi (k)$ have been introduced by

\begin{equation}
\varsigma =\dint\limits_{-\infty }^{\infty }d\eta p(\eta )(e^{\eta }-1),
\label{eq37}
\end{equation}

and

\begin{equation}
\xi (k)=\dint\limits_{-\infty }^{\infty }d\eta p(\eta )(e^{ik\eta }-1),
\label{eq38}
\end{equation}

with $p(\eta )$ being the probability density function of the magnitude of
asset price jumps.

It is easy to see that Eq.(\ref{eq36}) can be written in the form

\begin{equation}
C(S,T-t)=Se^{-q(T-t)}L_{1}(l)-Ke^{-r(T-t)}L_{2}(l),  \label{eq39}
\end{equation}

where functions $L_{1}(l)$ and $L_{2}(l)$ are defined as

\begin{equation}
L_{1}(l)=\frac{e^{-\lambda \varsigma (T-t)}}{2\pi }\dint\limits_{-\infty
}^{l}dz\dint\limits_{-\infty }^{\infty }dke^{ikz-z}\exp \left\{ \lambda
(T-t)\xi (k)\right\} ,  \label{eq40}
\end{equation}

\begin{equation}
L_{2}(l)=\frac{1}{2\pi }\dint\limits_{-\infty }^{l}dz\dint\limits_{-\infty
}^{\infty }dke^{ikz}\exp \left\{ \lambda (T-t)\xi (k)\right\} ,  \label{eq41}
\end{equation}

with $\varsigma $ and $\xi (k)$ given by Eqs.(\ref{eq37}) and (\ref{eq38}).

Thus, we found the new equation (\ref{eq39}) to value a European call option
when the stochastic dynamics of asset price is governed by Eq.(\ref{eq1}).

\subsubsection{Put option}

Substitution of Eq.(\ref{eq32}) into Eq.(\ref{eq26}) yields for the value of
a European put option

\begin{equation*}
P(x,T-t)=\frac{Ke^{-r(T-t)}}{2\pi }\dint\limits_{-\infty }^{\infty
}dx^{\prime }\dint\limits_{-\infty }^{\infty }dke^{ik(x-x^{\prime })}
\end{equation*}

\begin{equation}
\times \exp \left\{ [ik(r-q)+\lambda \dint\limits_{-\infty }^{\infty }d\eta
p(\eta )\left\{ e^{ik\eta }-1-ik(e^{\eta }-1)\right\} ](T-t)\right\}
\label{eq42}
\end{equation}

\begin{equation*}
\times \max (1-e^{x^{\prime }},0).
\end{equation*}

This equation can be presented as

\begin{equation}
P(S,T-t)=Ke^{-r(T-t)}\mathcal{L}_{2}(l)-Se^{-q(T-t)}\mathcal{L}_{1}(l),
\label{eq43}
\end{equation}

where $l$ is given by Eq.(\ref{eq35}) while functions $\mathcal{L}_{1}(l)$
and $\mathcal{L}_{2}(l)$ are introduced by

\begin{equation}
\mathcal{L}_{1}(l)=\frac{e^{-\lambda \varsigma (T-t)}}{2\pi }%
\dint\limits_{l}^{\infty }dz\dint\limits_{-\infty }^{\infty }dke^{ikz-z}\exp
\left\{ \lambda (T-t)\xi (k)\right\} ,  \label{eq44}
\end{equation}

\begin{equation}
\mathcal{L}_{2}(l)=\frac{1}{2\pi }\dint\limits_{l}^{\infty
}dz\dint\limits_{-\infty }^{\infty }dke^{ikz}\exp \left\{ \lambda (T-t)\xi
(k))\right\} ,  \label{eq45}
\end{equation}

with $\varsigma $ and $\xi (k)$ defined by Eqs.(\ref{eq37}) and (\ref{eq38}).

Thus, we obtained the new equation (\ref{eq43}) to value a European put
option when the stochastic dynamics of asset price is governed by Eq.(\ref%
{eq1}).

\section{Put-Call parity}

The put-call parity is a fundamental relationship between the values of
European call and put options, both with the same strike price $K$ and time
to expiry $T-t$. This relationship is a manifestation of the no-arbitrage
principle. The put-call parity equation is model independent and has the form

\begin{equation}
C(S,T-t)-P(S,T-t)=Se^{-q(T-t)}-Ke^{-r(T-t)}.  \label{eq46}
\end{equation}

To prove this relationship we use Eqs.(\ref{eq39}) and (\ref{eq43}) to obtain

\begin{equation}
C(S,T-t)-P(S,T-t)=  \label{eq47}
\end{equation}

\begin{equation*}
=Se^{-q(T-t)}(L_{1}(l)+\mathcal{L}_{1}(l))-Ke^{-r(T-t)}(L_{2}(l)+\mathcal{L}%
_{2}(l)),
\end{equation*}

From the definitions of functions $L$ given by Eqs.(\ref{eq40}) and (\ref%
{eq41}), and functions $\mathcal{L}$ given by Eqs.(\ref{eq44}) and (\ref%
{eq45}) we have

\begin{equation}
L_{1}(l)+\mathcal{L}_{1}(l)=1,  \label{eq48}
\end{equation}

and

\begin{equation}
L_{2}(l)+\mathcal{L}_{2}(l)=1.  \label{eq49}
\end{equation}

Let's show, for example, that equation (\ref{eq49}) holds. Indeed, with help
of Eqs.(\ref{eq41}) and (\ref{eq45}) we write for Eq.(\ref{eq49})

\begin{equation}
L_{2}(l)+\mathcal{L}_{2}(l)=\frac{1}{2\pi }{\LARGE (}\dint\limits_{-\infty
}^{\infty }dz\dint\limits_{-\infty }^{\infty }dke^{ikz}\exp \left\{ \lambda
(T-t)\xi (k)\right\} {\LARGE )}  \label{eq50}
\end{equation}

\begin{equation*}
=\dint\limits_{-\infty }^{\infty }dk\delta (k)\exp \left\{ \lambda (T-t)\xi
(k)\right\} =\exp \left\{ \lambda (T-t)\xi (0)\right\} =1,
\end{equation*}

here we used $\xi (k=0)=0$ as it follows immediately from Eq.(\ref{eq38}).

A similar consideration can be provided to prove Eq.(\ref{eq48}).

By substituting Eqs.(\ref{eq48}) and (\ref{eq49}) into Eq.(\ref{eq47}) we
complete the proof of put-call parity equation (\ref{eq46}).

\section{Greeks}

The purpose of this Section is to calculate the Greeks based on equation (%
\ref{eq39}).

In option pricing fundamentals and option trading, Greek letters are used to
define sensitivities (Greeks) of the option value in respect to a change in
either underlying price (i.e., asset price) or parameters (i.e., risk-free
rate, time to maturity, etc.). The Greeks can be considered as effective
tools to measure and manage the risk in an option position. The most common
Greeks are the first order derivatives: delta, rho, psi and theta as well as
gamma, a second-order derivative of the option value over underlying price.

To simplify calculations of the Greeks let's note that the following
equations hold

\begin{equation}
Se^{-q(T-t)}\frac{\partial L_{1}(l)}{\partial l}=Ke^{-r(T-t)}\frac{\partial
L_{2}(l)}{\partial l},  \label{eq51}
\end{equation}

\begin{equation}
Se^{-q(T-t)}\frac{\partial \mathcal{L}_{1}(l)}{\partial l}=Ke^{-r(T-t)}\frac{%
\partial \mathcal{L}_{2}(l)}{\partial l},  \label{eq52}
\end{equation}

where functions $L_{1}(l)$, $L_{2}(l)$, $\mathcal{L}_{1}(l)$, $\mathcal{L}%
_{2}(l)$ are given by Eqs.(\ref{eq40}), (\ref{eq41}), (\ref{eq44}) and (\ref%
{eq45}).

\subsection{Common Greeks}

The Greek delta for a call option $\Delta _{C}$ is given by

\begin{equation}
\Delta _{C}=\frac{\partial C(S,T-t)}{\partial S}=e^{-q(T-t)}L_{1}(l).
\label{eq53}
\end{equation}

The Greek rho $\rho _{C}$ for a call option is

\begin{equation}
\rho _{C}=\frac{\partial C(S,T-t)}{\partial r}=(T-t)Ke^{-r(T-t)}L_{2}(l).
\label{eq54}
\end{equation}

The Greek psi $\psi _{C}$ for a call option is the first partial derivative
of option value $C(S,T-t)$ with respect to the dividend rate $q$,

\begin{equation}
\psi _{C}=\frac{\partial C(S,T-t)}{\partial q}=-(T-t)Se^{-q(T-t)}L_{1}(l).
\label{eq55}
\end{equation}

The Greek theta $\Theta _{C}$ for a call option is defined by

\begin{equation*}
\Theta _{C}=-\frac{\partial C(S,T-t)}{\partial (T-t)}=\frac{\partial C(S,T-t)%
}{\partial t}.
\end{equation*}

Thus, we have

\begin{equation*}
\Theta _{C}=qSe^{-q(T-t)}L_{1}(l)-rKe^{-r(T-t)}L_{2}(l)+Se^{-q(T-t)}\lambda
\varsigma L_{1}(l)
\end{equation*}

\begin{equation}
-\lambda Se^{-q(T-t)}\frac{e^{-\lambda \varsigma (T-t)}}{2\pi }%
\dint\limits_{-\infty }^{l}dze^{-z}\dint\limits_{-\infty }^{\infty
}dke^{ikz}\xi (k)\exp \left\{ \lambda (T-t)\xi (k)\right\}  \label{eq56}
\end{equation}

\begin{equation*}
+\frac{\lambda Ke^{-r(T-t)}}{2\pi }\dint\limits_{-\infty
}^{l}dz\dint\limits_{-\infty }^{\infty }dke^{ikz}\xi (k)\exp \left\{ \lambda
(T-t)\xi (k)\right\} .
\end{equation*}

Finally, the second order sensitivity gamma $\Gamma _{C}$ for a call option
is

\begin{equation}
\Gamma _{C}=\frac{\partial ^{2}C(S,T-t)}{\partial S^{2}}=e^{-q(T-t)}\frac{%
\partial L_{1}(l)}{\partial l}\frac{\partial l}{\partial S}=\frac{e^{-q(T-t)}%
}{S}\frac{\partial L_{1}(l)}{\partial l}.  \label{eq57}
\end{equation}

Functions $L_{1}(l)$ and $L_{2}(l)$ in the formulas above are given by Eqs.(%
\ref{eq40}) and (\ref{eq41}).

Table 1 summarizes the common Greeks for a call option.

\begin{tabular}{|c|c|}
\hline
& {\small Call} \\ \hline
{\small Delta, }${\small \Delta }_{C}{\small =}\frac{\partial C(S,T-t)}{%
\partial S}$ & ${\small e}^{-q(T-t)}{\small L}_{1}{\small (l)}$ \\ \hline
{\small Gamma, }${\small \Gamma }_{C}{\small =}\frac{\partial C^{2}(S,T-t)}{%
\partial S^{2}}$ & $\frac{e^{-q(T-t)}}{S}\frac{\partial L_{1}(l)}{\partial l}%
=\frac{Ke^{-r(T-t)}}{S^{2}}\frac{\partial L_{2}(l)}{\partial l}$ \\ \hline
{\small Rho, }${\small \rho }_{C}{\small =}\frac{\partial C(S,T-t)}{\partial
r}$ & ${\small (T-t)Ke}^{-r(T-t)}{\small L}_{2}{\small (l)}$ \\ \hline
{\small Psi, }${\small \psi }_{C}{\small =}\frac{\partial C(S,T-t)}{\partial
q}$ & ${\small -(T-t)Se}^{-q(T-t)}{\small L}_{1}{\small (l)}$ \\ \hline
{\small Theta, }${\small \Theta }_{C}{\small =}\frac{\partial C(S,T-t)}{%
\partial t}$ & ${\small qSe}^{-q(T-t)}{\small L}_{1}{\small (l)-rKe}%
^{-r(T-t)}{\small L}_{2}{\small (l)}$ \\ 
& {\small +}$\lambda \varsigma Se^{-q(T-t)}L_{1}(l)$ \\ 
& ${\small -}\frac{{\small \lambda Se}^{-q(T-t)}e^{-\lambda \varsigma (T-t)}%
}{2\pi }\dint\limits_{-\infty }^{l}{\small dze}^{-z}\dint\limits_{-\infty
}^{\infty }{\small dke}^{ikz}{\small \xi (k)e}^{\lambda (T-t)\xi (k)}$ \\ 
& ${\small +}\frac{\lambda Ke^{-r(T-t)}}{2\pi }\dint\limits_{-\infty }^{l}%
{\small dz}\dint\limits_{-\infty }^{\infty }{\small dke}^{ikz}{\small \xi
(k)e}^{\lambda (T-t)\xi (k)}$ \\ \hline
\end{tabular}

Table 1.\textit{\ Common Greeks (Call option)}

Common Greeks for a put option can be easily found by using the put-call
parity equation (\ref{eq46}).

The Greek delta for a put option $\Delta _{P}$ is

\begin{equation}
\Delta _{P}=\frac{\partial P(S,T-t)}{\partial S}={\small -e}^{-q(T-t)}%
\mathcal{L}_{1}(l)=\Delta _{C}-e^{-q(T-t)},  \label{eq58}
\end{equation}

where $\Delta _{C}$ is the Greek delta for a call option given by Eq.(\ref%
{eq53}).

The Greek rho $\rho _{P}$ for a put option is

\begin{equation}
\rho _{P}=\frac{\partial P(S,T-t)}{\partial r}=-(T-t)Ke^{-r(T-t)}\mathcal{L}%
_{2}(l)=\rho _{C}-(T-t)Ke^{-r(T-t)},  \label{eq59}
\end{equation}

where $\rho _{C}$ is the Greek rho for a call option given by Eq.(\ref{eq54}%
).

The Greek psi $\psi _{P}$ for a put option is the first partial derivative
of option value $P(S,T-t)$ with respect to the dividend rate $q$,

\begin{equation}
\psi _{P}=\frac{\partial P(S,T-t)}{\partial q}=(T-t)Se^{-q(T-t)}\mathcal{L}%
_{1}(l)=\psi _{C}+(T-t)Se^{-q(T-t)},  \label{eq60}
\end{equation}

where $\psi _{C}$ is the Greek psi for a call option given by Eq.(\ref{eq55}%
).

The Greek theta $\Theta _{P}$ for a put option is defined by

\begin{equation*}
\Theta _{P}=-\frac{\partial P(S,T-t)}{\partial (T-t)}=\frac{\partial P(S,T-t)%
}{\partial t}.
\end{equation*}

Thus, from the put-call parity equation (\ref{eq46}) we have

\begin{equation}
\Theta _{P}=\Theta _{C}-qSe^{-q(T-t)}+rKe^{-r(T-t)},  \label{eq61}
\end{equation}

where $\Theta _{C}$ is the Greek theta for a call option given by Eq.(\ref%
{eq56}).

Finally, the second order sensitivity gamma $\Gamma _{P}$ for a put option is

\begin{equation}
\Gamma _{P}=\frac{\partial ^{2}P(S,T-t)}{\partial S^{2}}=-\frac{e^{-q(T-t)}}{%
S}\frac{\partial \mathcal{L}_{1}(l)}{\partial l}=\frac{e^{-q(T-t)}}{S}\frac{%
\partial L_{1}(l)}{\partial l}=\Gamma _{C},  \label{eq62}
\end{equation}

where $\Gamma _{C}$ is the Greek gamma for a call option given by Eq.(\ref%
{eq55}).

Hence, we conclude that

\begin{equation}
\Gamma _{P}=\Gamma _{C}.  \label{eq63}
\end{equation}

Functions $L_{1}(l)$, $\mathcal{L}_{1}(l)$, and $\mathcal{L}_{2}(l)$ in the
formulas above are given by Eqs.(\ref{eq40}), (\ref{eq44}) and (\ref{eq45}).

Table 2 summarizes the common Greeks for a put option.

\begin{tabular}{|c|c|}
\hline
& Put \\ \hline
{\small Delta, }${\small \Delta }_{P}{\small =}\frac{\partial P(S,T-t)}{%
\partial S}$ & ${\small -e}^{-q(T-t)}\mathcal{L}_{1}(l)=\Delta _{C}-{\small e%
}^{-q(T-t)}$ \\ \hline
{\small Gamma, }${\small \Gamma }_{P}{\small =}\frac{\partial P^{2}(S,T-t)}{%
\partial S^{2}}$ & -$\frac{e^{-q(T-t)}}{S}\frac{\partial \mathcal{L}_{1}(l)}{%
\partial l}=\frac{e^{-q(T-t)}}{S}\frac{\partial L_{1}(l)}{\partial l}=\Gamma
_{C}$ \\ \hline
{\small Rho, }${\small \rho }_{P}{\small =}\frac{\partial P(S,T-t)}{\partial
r}$ & $-(T-t)Ke^{-r(T-t)}\mathcal{L}_{2}(l)=\rho _{C}-(T-t)Ke^{-r(T-t)}$ \\ 
\hline
{\small Psi, }${\small \psi }_{P}{\small =}\frac{\partial P(S,T-t)}{\partial
q}$ & ${\small (T-t)Se}^{-q(T-t)}\mathcal{L}_{1}(l)=\psi _{C}+{\small (T-t)Se%
}^{-q(T-t)}$ \\ \hline
{\small Theta, }${\small \Theta }_{P}{\small =}\frac{\partial P(S,T-t)}{%
\partial t}$ & ${\small \Theta }_{C}{\small -qSe}^{-q(T-t)}{\small +rKe}%
^{-r(T-t)}$ \\ \hline
\end{tabular}

Table 2.\textit{\ Common Greeks (Put option)}

\subsection{New Greeks}

\subsubsection{Gaussian jumps model}

The new Greeks are option sensitivities associated with the market
parameters involved in the definition of shot noise process $F(t)$ (see, Eq.(%
\ref{eq2})). Besides parameter $\lambda $, which is the rate of asset price
jumps arrival, there can be other market parameters associated with
probability density function $p(\eta )$ involved into the definition of shot
noise $F(t)$. As soon as we specify the probability density function $p(\eta
)$, we will get the market parameters associated with it. Having these
parameters, we can introduce a few new Greeks.

At this point we assume, as an example, that the probability density
function of asset price jump magnitudes is a normal distribution

\begin{equation}
p(\eta )=\frac{1}{\sqrt{2\pi }\delta }\exp \{-\frac{(\eta -\nu )^{2}}{%
2\delta ^{2}}\},  \label{eq64}
\end{equation}

where the market parameters $\nu $ and $\delta ^{2}$ are the mean and
variance of asset price jump magnitudes.

The first order derivatives of a call option with respect to\ parameters $%
\lambda $, $\nu $ and $\delta $ will bring three new Greeks, which do not
exist in the Black-Scholes option pricing framework.

It follows immediately from put-call parity equation (\ref{eq46}) that all
three new Greeks are the same for call and put options. Thus, we need to
calculate these Greeks for call option only.

\subsubsection{Greek $\protect\kappa $}

We introduce the notation $\kappa _{C}$ for the derivative of call option
with respect to $\lambda $, which is the rate of asset price jumps arrival.
Hence, the definition of the Greek "kappa" $\kappa _{C}$ is

\begin{equation}
\kappa _{C}=\frac{\partial C(S,T-t)}{\partial \lambda }.  \label{eq65}
\end{equation}

The Greek $\kappa _{C}$ is a new Greek that does not exist in the
Black-Scholes framework, because of the absence of market parameter $\lambda 
$.

To find the Greek "kappa" we differentiate Eq.(\ref{eq39}) with respect to $%
\lambda $. The result is

\begin{equation*}
\frac{\kappa _{C}}{T-t}=-\varsigma Se^{-q(T-t)}L_{1}(l)
\end{equation*}

\begin{equation}
+Se^{-q(T-t)}\frac{e^{-\lambda \varsigma (T-t)}}{2\pi }\dint\limits_{-\infty
}^{l}dz\dint\limits_{-\infty }^{\infty }dke^{ikz-z}\xi (k)\exp \left\{
\lambda (T-t)\xi (k)\right\}  \label{eq66}
\end{equation}

\begin{equation*}
-\frac{Ke^{-r(T-t)}}{2\pi }\dint\limits_{-\infty
}^{l}dz\dint\limits_{-\infty }^{\infty }dke^{ikz}\xi (k)\exp \left\{ \lambda
(T-t)\xi (k)\right\} .
\end{equation*}

Comparing Eqs.(\ref{eq56}) and (\ref{eq66}) let's obtain the relationship
between $\Theta _{C}$ and $\kappa _{C}$

\begin{equation}
\kappa _{C}=\frac{T-t}{\lambda }%
(qSe^{-q(T-t)}L_{1}(l)-rKe^{-r(T-t)}L_{2}(l)-\Theta _{C}),  \label{eq67}
\end{equation}

or

\begin{equation}
\Theta _{C}=qSe^{-q(T-t)}L_{1}(l)-rKe^{-r(T-t)}L_{2}(l)-\frac{\lambda \kappa
_{C}}{T-t}.  \label{eq68}
\end{equation}

Thus, we discovered a new fundamental relationship between common Greek $%
\Theta _{C}$ and newly introduced Greek $\kappa _{C}$.

It has already been mentioned that the "kappa" for a put option $\kappa _{P}$
is the same as "kappa" for a call option $\kappa _{C}$,

\begin{equation}
\kappa _{P}=\kappa _{C},  \label{eq69}
\end{equation}

because of the put-call parity equation (\ref{eq46}).

The fundamental relationship between the common Greek $\Theta _{P}$ and
newly introduced $\kappa _{P}$ has the form

\begin{equation}
\Theta _{P}=-qSe^{-q(T-t)}\mathcal{L}_{1}(l)+rKe^{-r(T-t)}\mathcal{L}_{2}(l)-%
\frac{\lambda \kappa _{P}}{T-t},  \label{eq70}
\end{equation}

which can be easily verified with help of Eqs.(\ref{eq61}), (\ref{eq48}), (%
\ref{eq49}), (\ref{eq68}) and (\ref{eq69}).

Straightforward substitution of Eq.(\ref{eq39}) into the definition (\ref%
{eq65}) and calculation of the derivative of $C(S,T-t)$ with respect to $%
\lambda $ yield the identities for $\kappa _{C}$ with involvement of some
common Greeks. For example, it is easy to see that the identity holds

\begin{equation}
\kappa _{C}=S\frac{\partial \Delta _{C}}{\partial \lambda }-\frac{1}{T-t}%
\frac{\partial \rho _{C}}{\partial \lambda },  \label{eq71}
\end{equation}

where $\Delta _{C}$, and $\rho _{C}$ are defined by Eqs.(\ref{eq53}) and (%
\ref{eq54}).

\subsubsection{Greek $\protect\mu $}

Let us introduce the Greek "mu" $\mu _{C}$ as the first order derivative of
a call option with respect to parameter $\nu $, which is the mean of asset
price jump magnitudes,

\begin{equation}
\mu _{C}=\frac{\partial C(S,T-t)}{\partial \nu }.  \label{eq72}
\end{equation}

The Greek $\mu _{C}$ is a new Greek that does not exist in the Black-Scholes
framework, because of the absence of market parameter $\nu $.

Calculating the derivative of $C(S,T-t)$ with respect to $\nu $ yields

\begin{equation*}
\frac{\mu _{C}}{\lambda (T-t)}=-\frac{\partial \varsigma }{\partial \nu }%
Se^{-q(T-t)}L_{1}(l)
\end{equation*}

\begin{equation}
+Se^{-q(T-t)}\frac{e^{-\lambda \varsigma (T-t)}}{2\pi }\dint\limits_{-\infty
}^{l}dz\dint\limits_{-\infty }^{\infty }dke^{ikz-z}\frac{\partial \xi (k)}{%
\partial \nu }\exp \left\{ \lambda (T-t)\xi (k)\right\}  \label{eq73}
\end{equation}

\begin{equation*}
-\frac{Ke^{-r(T-t)}}{2\pi }\dint\limits_{-\infty
}^{l}dz\dint\limits_{-\infty }^{\infty }dke^{ikz}\frac{\partial \xi (k)}{%
\partial \nu }\exp \left\{ \lambda (T-t)\xi (k)\right\} .
\end{equation*}

It is easy to see from Eqs.(\ref{eq37}) and (\ref{eq38}) that for $p(\eta )$
given by Eq.(\ref{eq64}) we have

\begin{equation}
\frac{\partial \varsigma }{\partial \nu }=e^{\nu +\frac{\delta ^{2}}{2}%
}=(\varsigma +1),  \label{eq74}
\end{equation}

and

\begin{equation}
\frac{\partial \xi (k)}{\partial \nu }=e^{ik\nu -\frac{k^{2}\delta ^{2}}{2}%
}=ik(\xi (k)+1).  \label{eq75}
\end{equation}

Therefore, Eq.(\ref{eq73}) can be rewritten as

\begin{equation*}
\frac{\mu _{C}}{\lambda (T-t)}=-Se^{-q(T-t)}(\varsigma +1)L_{1}(l)
\end{equation*}

\begin{equation}
+Se^{-q(T-t)}\frac{e^{-\lambda \varsigma (T-t)}}{2\pi }\dint\limits_{-\infty
}^{l}dz\dint\limits_{-\infty }^{\infty }dke^{ikz-z}ik(\xi (k)+1)\exp \left\{
\lambda (T-t)\xi (k)\right\}  \label{eq76}
\end{equation}

\begin{equation*}
-\frac{Ke^{-r(T-t)}}{2\pi }\dint\limits_{-\infty
}^{l}dz\dint\limits_{-\infty }^{\infty }dke^{ikz}ik(\xi (k)+1)\exp \left\{
\lambda (T-t)\xi (k)\right\} .
\end{equation*}

Further, taking into account that

\begin{equation*}
ike^{ikz}=\frac{\partial }{\partial z}e^{ikz},
\end{equation*}

and performing integration over $dz$ we find

\begin{equation}
\frac{\mu _{C}}{\lambda (T-t)}=-\varsigma Se^{-q(T-t)}L_{1}(l)+  \label{eq77}
\end{equation}

\begin{equation*}
+Se^{-q(T-t)}\frac{e^{-\lambda \varsigma (T-t)}}{2\pi }\dint\limits_{-\infty
}^{l}dze^{-z}\dint\limits_{-\infty }^{\infty }dke^{ikz}\xi (k)\exp \left\{
\lambda (T-t)\xi (k)\right\} ,
\end{equation*}

which can be considered as the equation to calculate the Greek $\mu _{C}$.

Let us show that Eq.(\ref{eq77}) can be used to find new identities with
involvement of the Greek $\mu _{C}$ and some common Greeks. For instance, if
we rewrite Eq.(\ref{eq77}) in the form

\begin{equation}
\frac{\mu _{C}}{\lambda }=Se^{-q(T-t)}\left( \frac{\partial L_{1}(l)}{%
\partial \lambda }-\frac{\partial L_{1}(l)}{\partial l}\frac{\partial l}{%
\partial \lambda }\right) ,  \label{eq78}
\end{equation}

then it is easy to see that the identity holds

\begin{equation}
\frac{\mu _{C}}{\lambda }=S\frac{\partial \Delta _{C}}{\partial \lambda }%
+\varsigma (T-t)S^{2}\Gamma _{C}\ ,  \label{eq79}
\end{equation}

where the Greek $\Gamma _{C}$ is defined by Eq.(\ref{eq57}).

With the help of Eq.(\ref{eq71}) we can obtain another identity from Eq.(\ref%
{eq79})

\begin{equation}
\kappa _{C}=\frac{\mu _{C}}{\lambda }-\frac{1}{T-t}\frac{\partial \rho _{C}}{%
\partial \lambda }-\varsigma (T-t)S^{2}\Gamma _{C},  \label{eq80}
\end{equation}

which establishes the relationship between newly introduced Greeks $\kappa
_{C}$ and $\mu _{C}$.

\subsubsection{Greek $\protect\epsilon $}

A new Greek "epsilon" $\epsilon _{C}$\ is introduced as the derivative of a
call option with respect to parameter $\delta $, which is the standard
deviation of asset price jump magnitudes,

\begin{equation}
\epsilon _{C}=\frac{\partial C(S,T-t)}{\partial \delta }.  \label{eq81}
\end{equation}

The Greek $\epsilon _{C}$ is a new Greek that does not exist in the
Black-Scholes framework, because of the absence of market parameter $\delta $%
.

The derivative of $C(S,T-t)$ with respect to $\delta $ can be expressed as

\begin{equation*}
\frac{\epsilon _{C}}{\lambda (T-t)}=-\frac{\partial \varsigma }{\partial
\delta }Se^{-q(T-t)}L_{1}(l)
\end{equation*}

\begin{equation}
+Se^{-q(T-t)}\frac{e^{-\lambda \varsigma (T-t)}}{2\pi }\dint\limits_{-\infty
}^{l}dze^{-z}\dint\limits_{-\infty }^{\infty }dke^{ikz}\frac{\partial \xi (k)%
}{\partial \delta }\exp \left\{ \lambda (T-t)\xi (k)\right\}  \label{eq82}
\end{equation}

\begin{equation*}
-\frac{Ke^{-r(T-t)}}{2\pi }\dint\limits_{-\infty
}^{l}dz\dint\limits_{-\infty }^{\infty }dke^{ikz}\frac{\partial \xi (k)}{%
\partial \delta }\exp \left\{ \lambda (T-t)\xi (k)\right\} .
\end{equation*}

It follows from Eqs.(\ref{eq37}) and (\ref{eq38}) that

\begin{equation}
\frac{\partial \varsigma }{\partial \delta }=e^{\nu +\frac{\delta ^{2}}{2}%
}=\delta (\varsigma +1),  \label{eq83}
\end{equation}

and

\begin{equation}
\frac{\partial \xi (k)}{\partial \delta }=e^{ik\nu -\frac{k^{2}\delta ^{2}}{2%
}}=-k^{2}\delta (\xi (k)+1).  \label{eq84}
\end{equation}

Further, taking into account that

\begin{equation*}
-k^{2}e^{ikz}=\frac{\partial ^{2}}{\partial z^{2}}e^{ikz},
\end{equation*}

we have

\begin{equation*}
\frac{\epsilon _{C}}{\delta \lambda (T-t)}=-(\varsigma
+1)Se^{-q(T-t)}L_{1}(l)
\end{equation*}

\begin{equation*}
+Se^{-q(T-t)}\frac{e^{-\lambda \varsigma (T-t)}}{2\pi }\dint\limits_{-\infty
}^{l}dze^{-z}\dint\limits_{-\infty }^{\infty }dk\frac{\partial ^{2}}{%
\partial z^{2}}e^{ikz}(\xi (k)+1)\exp \left\{ \lambda (T-t)\xi (k)\right\}
\end{equation*}

\begin{equation*}
-\frac{Ke^{-r(T-t)}}{2\pi }\dint\limits_{-\infty
}^{l}dz\dint\limits_{-\infty }^{\infty }dk\frac{\partial ^{2}}{\partial z^{2}%
}e^{ikz}(\xi (k)+1)\exp \left\{ \lambda (T-t)\xi (k)\right\} .
\end{equation*}

Integration by parts over $dz$ yields

\begin{equation*}
\frac{\epsilon _{C}}{\lambda (T-t)\delta }=-\varsigma Se^{-q(T-t)}L_{1}(l)
\end{equation*}

\begin{equation}
+Se^{-q(T-t)}\frac{e^{-\lambda \varsigma (T-t)}}{2\pi }\dint\limits_{-\infty
}^{l}dze^{-z}\dint\limits_{-\infty }^{\infty }dke^{ikz}\xi (k)\exp \left\{
\lambda (T-t)\xi (k)\right\}  \label{eq85}
\end{equation}

\begin{equation*}
+\frac{Ke^{-r(T-t)}}{2\pi }\dint\limits_{-\infty }^{\infty }dke^{ikl}(\xi
(k)+1)\exp \left\{ \lambda (T-t)\xi (k)\right\} ,
\end{equation*}

which can be considered as the equation to calculate the Greek $\epsilon
_{C} $.

Comparing the equation above with Eq.(\ref{eq79}) let's conclude that the
following relationship holds

\begin{equation}
\frac{\epsilon _{C}}{\lambda (T-t)\delta }=\frac{\mu _{C}}{\lambda (T-t)}%
+S^{2}\Gamma _{C}  \label{eq86}
\end{equation}

\begin{equation*}
+\frac{Ke^{-r(T-t)}}{2\pi }\dint\limits_{-\infty }^{\infty }dke^{ikl}\xi
(k)\exp \left\{ \lambda (T-t)\xi (k)\right\} ,
\end{equation*}

where $\Gamma _{C}$ is given by Eq.(\ref{eq57}).

Straightforward substitution of Eq.(\ref{eq39}) into definition (\ref{eq81})
and calculation of the derivative of $C(S,T-t)$ with respect to $\delta $
yield the identities for $\epsilon _{C}$ with involvement of common Greeks $%
\Delta _{C}$ and $\rho _{C}$. For example, it is easy to see that the
identity holds

\begin{equation*}
\epsilon _{C}=S\frac{\partial \Delta _{C}}{\partial \delta }-\frac{1}{T-t}%
\frac{\partial \rho _{C}}{\partial \delta },
\end{equation*}

where $\Delta _{C}$, and $\rho _{C}$ are defined by Eqs.(\ref{eq53}) and (%
\ref{eq54}).

\section{Black-Scholes equation}

\subsection{Diffusion approximation}

We are aiming to show that the well-known Black-Scholes equation can be
obtained from the integro-differential option pricing equation (\ref{eq16}).
To get the Black-Scholes equation let's consider the market situation when
the variance of asset price jump magnitudes $\delta ^{2}\rightarrow 0$,
while the arrival rate of price jumps $\lambda \rightarrow \infty $ in such
a way that the product $\lambda \dint\limits_{-\infty }^{\infty }d\eta
p(\eta )\eta ^{2}$ remains finite. We call this case "diffusion
approximation".

Due to the condition $\delta ^{2}\rightarrow 0$, the expression under the
integral sign in Eq.(\ref{eq16}) can be expanded in $\eta $ up to the
second-order

\begin{equation}
\frac{\partial C(x,t)}{\partial t}+(r-q)\frac{\partial C(x,t)}{\partial x}
\label{eq87}
\end{equation}

\begin{equation*}
+\frac{\lambda }{2}\dint\limits_{-\infty }^{\infty }d\eta p(\eta )\eta
^{2}\left\{ \frac{\partial ^{2}C(x,t)}{\partial x^{2}}-\frac{\partial C(x,t)%
}{\partial x}\right\} -rC(x,t)=0.
\end{equation*}

To simplify the equations in downstream consideration it is convenient to
introduce the notation

\begin{equation}
\sigma ^{2}=\lambda \dint\limits_{-\infty }^{\infty }d\eta p(\eta )\eta ^{2}.
\label{eq88}
\end{equation}

In general, the "diffusion approximation" is the case when the mean of asset
price jump magnitudes $\nu \rightarrow 0$, the variance of asset price jump
magnitudes $\delta ^{2}\rightarrow 0$ and the arrival rate of price jumps $%
\lambda \rightarrow \infty $ while the products $\lambda
\dint\limits_{-\infty }^{\infty }d\eta p(\eta )\eta $ and $\lambda
\dint\limits_{-\infty }^{\infty }d\eta p(\eta )\eta ^{2}$ remain finite.

\subsection{Option pricing equation}

With help Eq.(\ref{eq88}), equation (\ref{eq87}) can be rewritten as

\begin{equation}
\frac{\partial C(x,t)}{\partial t}+\frac{\sigma ^{2}}{2}\frac{\partial
^{2}C(x,t)}{\partial x^{2}}+(r-q-\frac{\sigma ^{2}}{2})\frac{\partial C(x,t)%
}{\partial x}-rC(x,t)=0.  \label{eq89}
\end{equation}

Then the substitution

\begin{equation}
\frac{S}{K}=e^{x},  \label{eq90}
\end{equation}

yields

\begin{equation}
\frac{\partial C(S,t)}{\partial t}+\frac{\sigma ^{2}}{2}(S\frac{\partial }{%
\partial S})^{2}C(S,t)+(r-q-\frac{\sigma ^{2}}{2})S\frac{\partial C(S,t)}{%
\partial S}-rC(S,t)=0  \label{eq91}
\end{equation}

or

\begin{equation}
\frac{\partial C(S,t)}{\partial t}+\frac{\sigma ^{2}}{2}S^{2}\frac{\partial
^{2}C(S,t)}{\partial S^{2}}+(r-q)S\frac{\partial C(S,t)}{\partial S}%
-rC(S,t)=0.  \label{eq92}
\end{equation}

This is the famous Black-Scholes equation \cite{Black-Scholes}, which has to
be supplemented with the terminal condition given by Eq.(\ref{eq18}) for
call option and by Eq.(\ref{eq20}) for put option.

Thus, it has been shown that the new pricing equation (\ref{eq16}) goes into
the Black-Scholes equation (\ref{eq92}) in diffusion approximation. The
parameter $\sigma $ introduced by Eq.(\ref{eq88}) is called "volatility" of
asset price $S$.

If we go from time $t$ to the time to expiry $T-t$, then Eq.(\ref{eq92}) can
be rewritten as

\begin{equation}
-\frac{\partial C(S,T-t)}{\partial (T-t)}+\frac{\sigma ^{2}}{2}S^{2}\frac{%
\partial ^{2}C(S,T-t)}{\partial S^{2}}+(r-q)S\frac{\partial C(S,T-t)}{%
\partial S}-rC(S,T-t)=0,  \label{eq92T}
\end{equation}

while the terminal conditions (\ref{eq18}), (\ref{eq20}) became

\begin{equation}
C(S,T-t)|_{t=T}=C(S,0)=\max (S-K,0),  \label{eq93T}
\end{equation}

for a European call option, and

\begin{equation}
P(S,T-t)|_{t=T}=P(S,T)=\max (K-S,0).  \label{eq94T}
\end{equation}

for a European put option.

\subsection{The solution to Black-Scholes equation}

The straightforward solution to Eq.(\ref{eq92T}) with the terminal condition
(\ref{eq93T}) has been presented in Appendix A.

Here we show that solution (\ref{eq39}) to the integro-differential pricing
equation (\ref{eq23T}) goes into the Black-Scholes solution in diffusion
approximation.

In the diffusion approximation we have

\begin{equation}
\lambda \varsigma =\lambda \dint\limits_{-\infty }^{\infty }d\eta p(\eta
)(\eta +\frac{\eta ^{2}}{2})=\lambda \nu +\frac{\sigma ^{2}}{2},
\label{eq93}
\end{equation}

and

\begin{equation}
\lambda \xi (k)=\lambda \dint\limits_{-\infty }^{\infty }d\eta p(\eta
)(ik\eta -\frac{k^{2}\eta ^{2}}{2})=ik\lambda \nu -\frac{k^{2}\sigma ^{2}}{2}%
,  \label{eq94}
\end{equation}

where $\sigma ^{2}$ is given by Eq.(\ref{eq88}) and $\varsigma $ and $\xi
(k) $ are defined by Eqs.(\ref{eq37}) and (\ref{eq38}).

\subsubsection{Function $L_{1}(l)$ in diffusion approximation}

To find the equation for $L_{1}(l)$ in diffusion approximation we substitute
Eqs.(\ref{eq93}) and (\ref{eq94}) into Eq.(\ref{eq40})

\begin{equation}
L_{1}(l)\underset{\mathrm{diff}}{\rightarrow }L_{1}^{(\mathrm{diff})}=\frac{%
e^{-(\lambda \nu +\frac{\sigma ^{2}}{2})(T-t)}}{2\pi }\dint\limits_{-\infty
}^{x+(r-q-\lambda \nu -\frac{\sigma ^{2}}{2})(T-t)}dze^{-z}  \label{eq95}
\end{equation}

\begin{equation*}
\times \dint\limits_{-\infty }^{\infty }dke^{ikz}\exp \left\{ (ik\lambda \nu
-\frac{k^{2}\sigma ^{2}}{2})(T-t)\right\} .
\end{equation*}

Calculation of the integral over $dk$ results in

\begin{equation}
\frac{1}{2\pi }\dint\limits_{-\infty }^{\infty }dke^{ikz}\exp \left\{
(ik\lambda \nu -\frac{k^{2}\sigma ^{2}}{2})(T-t)\right\}  \label{eq96}
\end{equation}

\begin{equation*}
=\frac{1}{\sigma \sqrt{2\pi (T-t)}}\exp \left\{ -\frac{(z+\lambda \nu
(T-t))^{2}}{2\sigma ^{2}(T-t)}\right\} .
\end{equation*}

Substituting Eq.(\ref{eq96}) into Eq.(\ref{eq95}) and changing integration
variable $z\rightarrow y$

\begin{equation*}
z\rightarrow y=z+\lambda \nu (T-t),
\end{equation*}

yield

\begin{equation}
L_{1}^{(\mathrm{diff})}=\frac{1}{\sigma \sqrt{2\pi (T-t)}}  \label{eq97}
\end{equation}

\begin{equation*}
\times \dint\limits_{-\infty }^{x+(r-q-\frac{\sigma ^{2}}{2})(T-t)}dy\exp
\left\{ -\frac{1}{2}\left( \frac{y}{\sigma \sqrt{T-t}}+\sigma \sqrt{T-t}%
\right) ^{2}\right\} .
\end{equation*}

With help of a new integration variable $w$ 
\begin{equation*}
y\rightarrow w=\frac{y}{\sigma \sqrt{T-t}}+\sigma \sqrt{T-t},
\end{equation*}

we rewrite Eq.(\ref{eq97}) in the following way

\begin{equation}
L_{1}^{(\mathrm{diff})}=\frac{1}{\sqrt{2\pi }}\dint\limits_{-\infty }^{\frac{%
x+(r-q-\frac{\sigma ^{2}}{2})(T-t)}{\sigma \sqrt{T-t}}+\sigma \sqrt{T-t}%
}dw~e^{-\frac{w^{2}}{2}}.  \label{eq98}
\end{equation}

Then function $L_{1}^{(\mathrm{diff})}$ takes the form

\begin{equation}
L_{1}^{(\mathrm{diff})}=N(d_{1}),  \label{eq99}
\end{equation}

where function $N(d)$ is the cumulative distribution function of the
standard normal distribution \cite{AbramowitzStegun}

\begin{equation}
N(d)=\frac{1}{\sqrt{2\pi }}\dint\limits_{-\infty }^{d}dw~e^{-\frac{w^{2}}{2}%
},  \label{eq100}
\end{equation}

and parameter $d_{1}$ has been introduced by

\begin{equation}
d_{1}=\frac{x+(r-q-\frac{\sigma ^{2}}{2})(T-t)}{\sigma \sqrt{T-t}}+\sigma 
\sqrt{T-t}=\frac{x+(r-q+\frac{\sigma ^{2}}{2})(T-t)}{\sigma \sqrt{T-t}}.
\label{eq101}
\end{equation}

Thus, it has been shown that in diffusion approximation function $L_{1}(l)$
goes into function $N(d_{1})$,

\begin{equation}
L_{1}(l)\underset{\mathrm{diff}}{\rightarrow }L_{1}^{(\mathrm{diff}%
)}=N(d_{1}).  \label{eq102}
\end{equation}

The similar consideration brings the equation for function $\mathcal{L}%
_{1}(l)$ in diffusion approximation

\begin{equation}
\mathcal{L}_{1}(l)\underset{\mathrm{diff}}{\rightarrow }\mathcal{L}_{1}^{(%
\mathrm{diff})}=N(-d_{1}),  \label{eq103}
\end{equation}

with $N(d)$ and $d_{1}$ defined by Eqs.(\ref{eq100}) and (\ref{eq101}).

On a final note we present the equation for $\partial L_{1}(l)/\partial l$
in diffusion approximation

\begin{equation*}
\frac{\partial L_{1}(l)}{\partial l}\underset{\mathrm{diff}}{\rightarrow }%
\frac{1}{\sigma \sqrt{T-t}}N^{\prime }(d_{1}),
\end{equation*}

where $N^{\prime }(d)=\exp (-d^{2}/2)/\sqrt{2\pi }$ is derivative of $N(d)$
with respect to $d$ and $d_{1}$ is given by Eq.(\ref{eq101}).

\subsubsection{Function $L_{2}(l)$ in diffusion approximation}

To find the equation for $L_{2}(l)$ in diffusion approximation we substitute
Eqs.(\ref{eq93}) and (\ref{eq94}) into Eq.(\ref{eq41})

\begin{equation}
L_{2}(l)\underset{\mathrm{diff}}{\rightarrow }L_{2}^{(\mathrm{diff})}=\frac{1%
}{2\pi }\dint\limits_{-\infty }^{x+(r-q-\lambda \nu -\frac{\sigma ^{2}}{2}%
)(T-t)}dz  \label{eq104}
\end{equation}

\begin{equation*}
\times \dint\limits_{-\infty }^{\infty }dke^{ikz}\exp \left\{ (ik\lambda \nu
-\frac{k^{2}\sigma ^{2}}{2})(T-t)\right\} ,
\end{equation*}

Following the consideration provided while we calculated $L_{1}^{(\mathrm{%
diff})}$ we find

\begin{equation}
L_{2}^{(\mathrm{diff})}=N(d_{2}),  \label{eq105}
\end{equation}

where function $N(d)$ is defined by Eq.(\ref{eq100}) and $d_{2}$ has been
introduced by

\begin{equation}
d_{2}=\frac{x+(r-q-\frac{\sigma ^{2}}{2})(T-t)}{\sigma \sqrt{T-t}}%
=d_{1}-\sigma \sqrt{T-t}.  \label{eq106}
\end{equation}

Thus, it has been shown that in diffusion approximation function $L_{2}(l)$
goes into function $N(d_{2})$.

\begin{equation*}
L_{2}(l)\underset{\mathrm{diff}}{\rightarrow }L_{2}^{(\mathrm{diff}%
)}=N(d_{2}).
\end{equation*}

In a similar way the equation for $\mathcal{L}_{2}(l)$ can be obtained

\begin{equation}
\mathcal{L}_{2}(l)\underset{\mathrm{diff}}{\rightarrow }\mathcal{L}_{2}^{(%
\mathrm{diff})}=N(-d_{2}),  \label{eq107}
\end{equation}

with $N(d)$ and $d_{2}$ defined by Eqs.(\ref{eq100}) and (\ref{eq106}).

On a final note we present the equation for $\partial L_{2}(l)/\partial l$
in diffusion approximation

\begin{equation*}
\frac{\partial L_{2}(l)}{\partial l}\underset{\mathrm{diff}}{\rightarrow }%
\frac{1}{\sigma \sqrt{T-t}}N^{\prime }(d_{2}),
\end{equation*}

where $N^{\prime }(d)=\exp (-d^{2}/2)/\sqrt{2\pi }$ is derivative of $N(d)$
with respect to $d$, and $d_{2}$ is given by Eq.(\ref{eq106}).

\subsubsection{Solution to option pricing equation in diffusion approximation%
}

Using Eqs.(\ref{eq99}) and (\ref{eq105}) we can write Eq.(\ref{eq39}) in
diffusion approximation in the form

\begin{equation}
C_{BS}(S,T-t)=Se^{-q(T-t)}N(d_{1})-Ke^{-r(T-t)}N(d_{2}),  \label{eq108}
\end{equation}

where $C_{BS}(S,T-t)$ stands for the value of a European call option in the
Black-Scholes model.

Using Eqs.(\ref{eq103}) and (\ref{eq107}) we can write Eq.(\ref{eq43}) in
diffusion approximation in the form

\begin{equation}
P_{BS}(S,T-t)=Ke^{-r(T-t)}N(-d_{2})-Se^{-q(T-t)}N(-d_{1}),  \label{eq109}
\end{equation}

where $P_{BS}(S,T-t)$ stands for the value of a European put option in the
Black-Scholes model.

Hence, we see that in diffusion approximation, solutions (\ref{eq39}) and (%
\ref{eq43}) go into the Black-Scholes solutions for call and put options.
The key feature of diffusion approximation is the Black-Scholes volatility
defined by Eq.(\ref{eq88}). In other words, Eq.(\ref{eq88}) sheds light into
the stochastic dynamic origin of the Black-Scholes volatility, which emerges
naturally in the diffusion approximation.

\subsection{Greeks in diffusion approximation}

\subsubsection{Greek vega}

In the Black-Scholes world there is a volatility $\sigma $. Hence, we can
consider the derivative of call and put options with respect to $\sigma $.
This is the Greek "vega"\footnote{%
Actually, there is no such Greek name as vega and $\upsilon $ does not
belong to Greek symbols.
\par
{}
\par
{}}.

For instance, for a European call option we have vega $\upsilon _{C}$ which
is

\begin{equation}
\upsilon _{C}=\frac{\partial C(S,T-t)}{\partial \sigma }=Se^{-q(T-t)}\sqrt{%
T-t}N^{\prime }(d_{1})=Ke^{-r(T-t)}\sqrt{T-t}N^{\prime }(d_{2}),
\label{eq110}
\end{equation}

where $N^{\prime }(d)$ stands for derivative of $N(d)$ with respect to $d$,

\begin{equation}
N^{\prime }(d)=\frac{1}{\sqrt{2\pi }}\exp (-d^{2}/2),  \label{eq111}
\end{equation}

parameters $d_{1}$ and $d_{2}$ are defined by Eqs.(\ref{eq101}) and (\ref%
{eq106}).

It follows from the put-call parity law (\ref{eq46}), which holds for the
Black-Scholes solutions, that

\begin{equation}
\upsilon _{P}=\frac{\partial P(S,T-t)}{\partial \sigma }=\frac{\partial
C(S,T-t)}{\partial \sigma }=\upsilon _{C},  \label{eq112}
\end{equation}

that is, the Greeks vega for call and put options are the same.

\subsubsection{Black-Scholes Greeks}

It follows from Eqs.(\ref{eq102}), (\ref{eq103}), (\ref{eq105}) and (\ref%
{eq107}) that in diffusion approximation all common Greeks presented in
Tables 1 and 2 go into the well-known Black-Scholes Greeks.

Table 3 summarizes the Black-Scholes Greeks for a call option.

\begin{tabular}{|c|c|}
\hline
& {\small Call (Black-Scholes pricing equation)} \\ \hline
{\small Delta, }${\small \Delta }_{C}{\small =}\frac{\partial C(S,T-t)}{%
\partial S}$ & ${\small e}^{-q(T-t)}N(d_{1}{\small )}$ \\ \hline
{\small Gamma, }${\small \Gamma }_{C}{\small =}\frac{\partial C^{2}(S,T-t)}{%
\partial S^{2}}$ & $\frac{e^{-q(T-t)}}{S\sigma \sqrt{T-t}}N^{\prime }(d_{1}%
{\small )}=\frac{Ke^{-r(T-t)}}{S^{2}\sigma \sqrt{T-t}}N^{\prime }(d_{2}%
{\small )}$ \\ \hline
{\small Rho, }${\small \rho }_{C}{\small =}\frac{\partial C(S,T-t)}{\partial
r}$ & ${\small (T-t)Ke}^{-r(T-t)}N(d_{2}{\small )}$ \\ \hline
{\small Psi, }${\small \psi }_{C}{\small =}\frac{\partial C(S,T-t)}{\partial
q}$ & ${\small -(T-t)Se}^{-q(T-t)}N(d_{1}{\small )}$ \\ \hline
{\small Theta, }${\small \Theta }_{C}{\small =}\frac{\partial C(S,T-t)}{%
\partial t}$ & $qSe^{-q(T-t)}N(d_{1})-rKe^{-r(T-t)}N(d_{2})-\frac{{\small %
\sigma S}e^{-q(T-t)}N^{\prime }(d_{1}{\small )}}{2\sqrt{T-t}}$ \\ \hline
Vega ${\small \upsilon }_{C}{\small =}\frac{\partial C(S,T-t)}{\partial
\sigma }$ & ${\small Se}^{-q(T-t)}\sqrt{T-t}{\small N}^{\prime }{\small (d}%
_{1}{\small )=Ke}^{-r(T-t)}\sqrt{T-t}{\small N}^{\prime }{\small (d}_{2}%
{\small )}$ \\ \hline
\end{tabular}

Table 3.\textit{\ Black-Scholes Greeks (Call option)}

\section{Generalized option pricing framework}

\subsection{Geometric Brownian motion and Geometric Shot Noise}

The new option pricing framework presented in Sec.3 can be easily
generalized to accommodate a superposition of Geometric Brownian motion and
a Geometric Shot Noise motion. Indeed, if besides the Geometric Shot Noise
motion we have the Geometric Brownian motion, then Eq.(\ref{eq1}) becomes

\begin{equation}
dS=\mu Sdt+\sigma SdW+SF(t)dt,  \label{eq113}
\end{equation}

where $\mu $ and $\sigma $ are constants belonging to the Brownian motion
process, $W$ is a standard Wiener process, and the shot noise process $F(t)$
has been introduced by Eq.(\ref{eq2}).

\subsection{Generalized arbitrage-free integro-differential option pricing
equation}

If asset price follows Eq.(\ref{eq113}) then the generalized arbitrage-free
integro-differential option pricing equation has a form

\begin{equation}
\frac{\partial C(x,t)}{\partial t}+\frac{\sigma ^{2}}{2}\frac{\partial
^{2}C(x,t)}{\partial x^{2}}+(r-q-\frac{\sigma ^{2}}{2})\frac{\partial C(x,t)%
}{\partial x}  \label{eq114}
\end{equation}

\begin{equation*}
+\lambda \dint\limits_{-\infty }^{\infty }d\eta p(\eta )\left\{ C(x+\eta
,t)-C(x,t)-(e^{\eta }-1)\frac{\partial C(x,t)}{\partial x}\right\}
-rC(x,t)=0,
\end{equation*}

here $x$ is given by Eq.(\ref{eq17}), $C(x,t)$ is the value of a European
call option on divident-paying asset, $S$ is asset price governed by Eq.(\ref%
{eq113}), $K$ is the strike price, $r$ is the risk-free interest rate, $q$
is continuously paid dividend yield, which is a constant, $p(\eta )$ is the
probability density function involved into Eq.(\ref{eq2}).

If we go from $C(x,t)$ to $C(S,t)$ where $x$ and $S$ are related to each
other by Eq.(\ref{eq17}), then we can write the generalized option pricing
equation in the form

\begin{equation}
\frac{\partial C(S,t)}{\partial t}+\frac{\sigma ^{2}}{2}S^{2}\frac{\partial
^{2}C(S,t)}{\partial S^{2}}+(r-q)S\frac{\partial C(S,t)}{\partial S}
\label{eq115}
\end{equation}

\begin{equation*}
+\lambda \dint\limits_{-\infty }^{\infty }d\eta p(\eta )\left\{ C(Se^{\eta
},t)-C(S,t)-(e^{\eta }-1)S\frac{\partial C(S,t)}{\partial S}\right\}
-rC(S,t)=0,
\end{equation*}

with the terminal condition given either by Eq.(\ref{eq18}) or by Eq.(\ref%
{eq20}).

In the case when $\lambda =0$, Eq.(\ref{eq115}) becomes the Black-Scholes
equation (see, Eq.(\ref{eq92})). In the case when $\sigma =0$, Eq.(\ref%
{eq115}) becomes the integro-differential option pricing equation (see, Eq.(%
\ref{eq22})).

If we go from time $t$ to the time to expiry $T-t$, then Eq.(\ref{eq115})
can be rewritten as

\begin{equation*}
-\frac{\partial C(S,T-t)}{\partial (T-t)}+\frac{\sigma ^{2}}{2}S^{2}\frac{%
\partial ^{2}C(S,T-t)}{\partial S^{2}}+(r-q)S\frac{\partial C(S,T-t)}{%
\partial S}
\end{equation*}

\begin{equation}
+\lambda \dint\limits_{-\infty }^{\infty }d\eta p(\eta )\left\{ C(Se^{\eta
},T-t)-C(S,T-t)-(e^{\eta }-1)S\frac{\partial C(S,T-t)}{\partial S}\right\}
\label{eq115T}
\end{equation}

\begin{equation*}
-rC(S,T-t)=0,
\end{equation*}

with the terminal condition (\ref{eq93T}) for a European call option and the
terminal condition (\ref{eq94T}) for a European put option.

The solution to Eq.(\ref{eq115T}) with the terminal condition (\ref{eq93T})
is

\begin{equation}
C(S,T-t)=Se^{-q(T-t)}\mathbf{L}_{1}(l_{\sigma })-Ke^{-r(T-t)}\mathbf{L}%
_{2}(l_{\sigma }),  \label{eq116}
\end{equation}

where new functions $\mathbf{L}_{1}(l_{\sigma })$ and $\mathbf{L}%
_{2}(l_{\sigma })$ are introduced by

\begin{equation}
\mathbf{L}_{1}(l_{\sigma })=\frac{e^{(-\frac{\sigma ^{2}}{2}-\lambda
\varsigma )(T-t)}}{2\pi }\dint\limits_{-\infty }^{l_{\sigma
}}dz\dint\limits_{-\infty }^{\infty }dke^{ikz-z}\exp \left\{ [-\frac{\sigma
^{2}k^{2}}{2}+\lambda \xi (k)](T-t)\right\} ,  \label{eq117}
\end{equation}

and

\begin{equation}
\mathbf{L}_{2}(l_{\sigma })=\frac{1}{2\pi }\dint\limits_{-\infty
}^{l_{\sigma }}dz\dint\limits_{-\infty }^{\infty }dke^{ikz}\exp \left\{ [-%
\frac{\sigma ^{2}k^{2}}{2}+\lambda \xi (k)](T-t)\right\} ,  \label{eq118}
\end{equation}

with $\varsigma $ and $\xi (k)$ given by Eqs.(\ref{eq37}) and (\ref{eq38}),
and parameter $l_{\sigma }$ defined by

\begin{equation}
l_{\sigma }=x+(r-q-\frac{\sigma ^{2}}{2}-\lambda \varsigma )(T-t).
\label{eq119}
\end{equation}

Generalized option pricing formula (\ref{eq116}) for a European call option
follows from asset price stochastic dynamics (\ref{eq113}) modeled by
superposition of Geometric Brownian motion and a Geometric Shot Noise.

Let us consider two limit cases of Eq.(\ref{eq116}).

1. In the limit case when $\sigma =0$ we have

\begin{equation}
\mathbf{L}_{1}(l_{\sigma })|_{\sigma =0}=L_{1}(l),  \label{eq120}
\end{equation}

\begin{equation}
\mathbf{L}_{2}(l_{\sigma })|_{\sigma =0}=L_{2}(l),  \label{eq121}
\end{equation}

and Eq.(\ref{eq116}) goes into the option pricing equation (\ref{eq39}) with 
$l$ defined by Eq.(\ref{eq35}).

2. In the limit case when $\lambda =0$ we have

\begin{equation}
\mathbf{L}_{1}(l_{\sigma })|_{\lambda =0}=N(d_{1}),  \label{eq122}
\end{equation}

\begin{equation}
\mathbf{L}_{2}(l_{\sigma })|_{\lambda =0}=N(d_{2}),  \label{eq123}
\end{equation}

and Eq.(\ref{eq116}) goes into the Black-Scholes option pricing formula (\ref%
{eq108}) with $d_{1}$ and $d_{2}$ defined by Eqs.(\ref{eq101}) and (\ref%
{eq106}).

Hence, Eq.(\ref{eq116}) describes the impact of the interplay between
Gaussian Geometric Brownian motion and non-Gaussian Geometric Shot Noise on
the value of a European call option.

On a final note, let us present the generalized option pricing formula for a
European put option. The solution to Eq.(\ref{eq115T}) with terminal
condition (\ref{eq94T}) is

\begin{equation}
P(S,T-t)=Ke^{-r(T-t)}\mathfrak{L}_{2}(l_{\sigma })-Se^{-q(T-t)}\mathfrak{L}%
_{1}(l_{\sigma }),  \label{eq124}
\end{equation}

where new functions $\mathfrak{L}_{1}(l)$ and $\mathfrak{L}_{2}(l)$ are
introduced by

\begin{equation}
\mathfrak{L}_{1}(l_{\sigma })=\frac{e^{(-\frac{\sigma ^{2}}{2}-\lambda
\varsigma )(T-t)}}{2\pi }\dint\limits_{l_{\sigma }}^{\infty
}dz\dint\limits_{-\infty }^{\infty }dke^{ikz-z}\exp \left\{ [-\frac{\sigma
^{2}k^{2}}{2}+\lambda \xi (k)](T-t)\right\} ,  \label{eq125}
\end{equation}

and

\begin{equation}
\mathfrak{L}_{2}(l_{\sigma })=\frac{1}{2\pi }\dint\limits_{l_{\sigma
}}^{\infty }dz\dint\limits_{-\infty }^{\infty }dke^{ikz}\exp \left\{ [-\frac{%
\sigma ^{2}k^{2}}{2}+\lambda \xi (k)](T-t)\right\} ,  \label{eq126}
\end{equation}

and the parameter $l_{\sigma }$ is defined by Eq.(\ref{eq119}).

Let us consider two limit cases of Eq.(\ref{eq124}).

1. In the limit case when $\sigma =0$ we have

\begin{equation}
\mathfrak{L}_{1}(l_{\sigma })|_{\sigma =0}=\mathcal{L}_{1}(l),  \label{eq127}
\end{equation}

\begin{equation}
\mathfrak{L}_{2}(l_{\sigma })|_{\sigma =0}=\mathcal{L}_{2}(l),  \label{eq128}
\end{equation}

and Eq.(\ref{eq124}) goes into the option pricing equation (\ref{eq43}) with 
$l$ defined by Eq.(\ref{eq35}).

2. In the limit case when $\lambda =0$ we have

\begin{equation}
\mathfrak{L}_{1}(l_{\sigma })|_{\lambda =0}=N(-d_{1}),  \label{eq129}
\end{equation}

\begin{equation}
\mathfrak{L}_{2}(l_{\sigma })|_{\lambda =0}=N(-d_{2}),  \label{eq130}
\end{equation}

and Eq.(\ref{eq124}) goes into the Balck-Scholes option pricing formula (\ref%
{eq109}) with $d_{1}$ and $d_{2}$ defined by Eqs.(\ref{eq101}) and (\ref%
{eq106}).

Hence, Eq.(\ref{eq124}) describes the impact of the interplay between
Gaussian Geometric Brownian motion and non-Gaussian Geometric Shot Noise on
the value of a European put option.

It is easy to see that the value of call option given by Eq.(\ref{eq116})
and the value of put option given by Eq.(\ref{eq124}) satisfy the
fundamental put-call relationship (\ref{eq46}).

\subsection{Generalized Greeks}

Here we present the equations for common generalized Greeks in the case when
the value of a European call option is given by Eq.(\ref{eq116}).

The generalized Greek delta for a call option $\Delta _{C}$ is

\begin{equation}
\Delta _{C}=\frac{\partial C(S,T-t)}{\partial S}=e^{-q(T-t)}\mathbf{L}%
_{1}(l_{\sigma }).  \label{eqG1}
\end{equation}

The generalized Greek rho $\rho _{C}$ for a call option is

\begin{equation}
\rho _{C}=\frac{\partial C(S,T-t)}{\partial r}=(T-t)Ke^{-r(T-t)}\mathbf{L}%
_{2}(l_{\sigma }).  \label{eqG2}
\end{equation}

The generalized Greek psi $\psi _{C}$ for a call option is

\begin{equation}
\psi _{C}=\frac{\partial C(S,T-t)}{\partial q}=-(T-t)Se^{-q(T-t)}\mathbf{L}%
_{1}(l_{\sigma }).  \label{eqG3}
\end{equation}

The generalized Greek theta $\Theta _{C}$ for a call option is

\begin{equation*}
\Theta _{C}=\frac{\partial C(S,T-t)}{\partial t}
\end{equation*}

\begin{equation*}
=qSe^{-q(T-t)}\mathbf{L}_{1}(l_{\sigma })-rKe^{-r(T-t)}\mathbf{L}%
_{2}(l_{\sigma })+Se^{-q(T-t)}(\frac{\sigma ^{2}}{2}+\lambda \varsigma )%
\mathbf{L}_{1}(l_{\sigma })
\end{equation*}

\begin{equation}
-Se^{-q(T-t)}\frac{e^{(-\frac{\sigma ^{2}}{2}-\lambda \varsigma )(T-t)}}{%
2\pi }  \label{eqG4}
\end{equation}

\begin{equation*}
\times \dint\limits_{-\infty }^{l}dze^{-z}\dint\limits_{-\infty }^{\infty
}dke^{ikz}(-\frac{\sigma ^{2}k^{2}}{2}+\lambda \xi (k))\exp \left\{ [-\frac{%
\sigma ^{2}k^{2}}{2}+\lambda \xi (k)](T-t)\right\}
\end{equation*}

\begin{equation*}
+\frac{\lambda Ke^{-r(T-t)}}{2\pi }\dint\limits_{-\infty
}^{l}dz\dint\limits_{-\infty }^{\infty }dke^{ikz}(-\frac{\sigma ^{2}k^{2}}{2}%
+\lambda \xi (k))\exp \left\{ [-\frac{\sigma ^{2}k^{2}}{2}+\lambda \xi
(k)](T-t)\right\} .
\end{equation*}

The second order sensitivity generalized gamma $\Gamma _{C}$ for a call
option is

\begin{equation}
\Gamma _{C}=\frac{\partial ^{2}C(S,T-t)}{\partial S^{2}}=e^{-q(T-t)}\frac{%
\partial \mathbf{L}_{1}(l_{\sigma })}{\partial l_{\sigma }}\frac{\partial
l_{\sigma }}{\partial S}=\frac{e^{-q(T-t)}}{S}\frac{\partial \mathbf{L}%
_{1}(l_{\sigma })}{\partial l_{\sigma }}.  \label{eqG5}
\end{equation}

Finally, the generalized Greek vega $\upsilon _{C}$ for a call option is

\begin{equation*}
\upsilon _{C}=\frac{\partial C(S,T-t)}{\partial \sigma }=-\sigma
(T-t)Se^{-q(T-t)}\mathbf{L}_{1}(l_{\sigma })
\end{equation*}

\begin{equation}
-\frac{Se^{-q(T-t)}\sigma (T-t)e^{(-\frac{\sigma ^{2}}{2}-\lambda \varsigma
)(T-t)}}{2\pi }  \label{eqG6}
\end{equation}

\begin{equation*}
\times \dint\limits_{-\infty }^{l_{\sigma }}dz\dint\limits_{-\infty
}^{\infty }dke^{ikz-z}k^{2}\exp \left\{ [-\frac{\sigma ^{2}k^{2}}{2}+\lambda
\xi (k)](T-t)\right\}
\end{equation*}%
\begin{equation*}
+\frac{Ke^{-r(T-t)}\sigma (T-t)}{2\pi }\dint\limits_{-\infty }^{l_{\sigma
}}dz\dint\limits_{-\infty }^{\infty }dke^{ikz}k^{2}\exp \left\{ [-\frac{%
\sigma ^{2}k^{2}}{2}+\lambda \xi (k)](T-t)\right\} .
\end{equation*}

Functions $\mathbf{L}_{1}(l_{\sigma })$ and $\mathbf{L}_{2}(l_{\sigma })$ in
the formulas above are given by Eqs.(\ref{eq117}) and (\ref{eq118}).

Table 4 summarizes the common generalized Greeks for a call option.

\begin{tabular}{|c|c|}
\hline
& {\small Call} \\ \hline
{\small Delta, }${\small \Delta }_{C}{\small =}\frac{\partial C(S,T-t)}{%
\partial S}$ & ${\small e}^{-q(T-t)}\mathbf{L}_{1}(l_{\sigma })$ \\ \hline
{\small Gamma, }${\small \Gamma }_{C}{\small =}\frac{\partial C^{2}(S,T-t)}{%
\partial S^{2}}$ & $\frac{e^{-q(T-t)}}{S}\frac{\partial \mathbf{L}%
_{1}(l_{\sigma })}{\partial l_{\sigma }}=\frac{Ke^{-r(T-t)}}{S^{2}}\frac{%
\partial \mathbf{L}_{2}(l_{\sigma })}{\partial l_{\sigma }}$ \\ \hline
{\small Rho, }${\small \rho }_{C}{\small =}\frac{\partial C(S,T-t)}{\partial
r}$ & ${\small (T-t)Ke}^{-r(T-t)}\mathbf{L}_{2}(l_{\sigma })$ \\ \hline
{\small Psi, }${\small \psi }_{C}{\small =}\frac{\partial C(S,T-t)}{\partial
q}$ & ${\small -(T-t)Se}^{-q(T-t)}\mathbf{L}_{1}(l_{\sigma })$ \\ \hline
{\small Theta, }${\small \Theta }_{C}{\small =}\frac{\partial C(S,T-t)}{%
\partial t}$ & ${\small qSe}^{-q(T-t)}\mathbf{L}_{1}(l_{\sigma }){\small -rKe%
}^{-r(T-t)}\mathbf{L}_{2}(l_{\sigma })$ \\ 
& {\small +(}$\sigma ^{2}/2+\lambda \varsigma )Se^{-q(T-t)}\mathbf{L}%
_{1}(l_{\sigma })$ \\ 
& ${\small -}\frac{{\small Se}^{-q(T-t)}e^{(-\frac{\sigma ^{2}}{2}-\lambda
\varsigma )(T-t)}}{2\pi }\dint\limits_{-\infty }^{l}{\small dze}^{-z}$ \\ 
& $\times \dint\limits_{-\infty }^{\infty }{\small dke}^{ikz}{\small (-}%
\frac{\sigma ^{2}k^{2}}{2}{\small +\lambda \xi (k))e}^{\{[-\frac{\sigma
^{2}k^{2}}{2}+\lambda \xi (k)](T-t)\}}$ \\ 
& ${\small +}\frac{\lambda Ke^{-r(T-t)}}{2\pi }\dint\limits_{-\infty }^{l}%
{\small dz}$ \\ 
& $\times \dint\limits_{-\infty }^{\infty }{\small dke}^{ikz}{\small (-}%
\frac{\sigma ^{2}k^{2}}{2}{\small +\lambda \xi (k))e}^{\{[-\frac{\sigma
^{2}k^{2}}{2}+\lambda \xi (k)](T-t)\}}$ \\ \hline
{\small Vega, }$\upsilon _{C}=\frac{\partial C(S,T-t)}{\partial \sigma }$ & $%
{\small -\sigma (T-t)Se}^{-q(T-t)}\mathbf{L}_{1}(l_{\sigma })$ \\ 
& $-\frac{Se^{-q(T-t)}\sigma (T-t)e^{(-\frac{\sigma ^{2}}{2}-\lambda
\varsigma )(T-t)}}{2\pi }\dint\limits_{-\infty }^{l_{\sigma }}dz$ \\ 
& $\times \dint\limits_{-\infty }^{\infty }dke^{ikz-z}k^{2}{\small e}^{\{[-%
\frac{\sigma ^{2}k^{2}}{2}+\lambda \xi (k)](T-t)\}}$ \\ 
& $+\frac{Ke^{-r(T-t)}\sigma (T-t)}{2\pi }\dint\limits_{-\infty }^{l_{\sigma
}}dz$ \\ 
& $\times \dint\limits_{-\infty }^{\infty }dke^{ikz}k^{2}{\small e}^{\{[-%
\frac{\sigma ^{2}k^{2}}{2}+\lambda \xi (k)](T-t)\}}$ \\ \hline
\end{tabular}

Table 4.\textit{\ Common generalized Greeks (Call option)}

Let us consider two limit cases of Eqs.(\ref{eqG1})-(\ref{eqG6}).

1. In the limit case when $\sigma =0$ the generalized Greek vega defined by
Eq.(\ref{eqG6}) does not exist, while the generalized Greeks defined by Eqs.(%
\ref{eqG1})-(\ref{eqG5}) go into the common Greeks introduced by Eqs.(\ref%
{eq53})-(\ref{eq57}) due to Eqs.(\ref{eq120}) and (\ref{eq121}).

2. In the limit case when $\lambda =0$ the generalized Greeks defined by
Eqs.(\ref{eqG1})-(\ref{eqG6}) go into the well known Black-Scholes Greeks
due to Eqs.(\ref{eq122}) and (\ref{eq123}). The Black-Scholes Greeks are
presented in Table 3.

The common generalized Greeks for a put option can be found by using the
put-call parity equation (\ref{eq46}).

Hence, new equations (\ref{eqG1})-(\ref{eqG6}) describe the impact of
interplay between Gaussian Geometric Brownian motion and non-Gaussian
Geometric Shot Noise on the common generalized Greeks.

\section{Short term interest rate dynamics}

\subsection{Langevin equation with shot noise}

We introduce a new stochastic model to describe short term interest rate
dynamics,

\begin{equation}
dr=-ardt+F(t)dt,  \label{eq131}
\end{equation}

which is a stochastic differential equation of the Langevin type with the
shot noise $F(t)$ given by Eq.(\ref{eq2}) and the parameter $a$ ($a>0$)
being the speed at which $r(t)$ goes to its equilibrium level at $%
t\rightarrow \infty .$

The short term interest rate $r(t)$ governed by Eq.(\ref{eq131}) is a
non-Gaussian random process.

It is easy to see from Eq.(\ref{eq131}) that if at the time moment $t$ the
short term interest rate is $r(t)$, then later on, at the time moment $s$, $%
s>t$,\ it will be

\begin{equation}
r(s)=e^{-a(s-t)}r(t)+\dint\limits_{t}^{s}d\tau e^{-a(s-\tau )}F(\tau )
\label{eq132}
\end{equation}

or

\begin{equation}
r(s)=e^{-as}r_{0}+\dint\limits_{0}^{s}d\tau e^{-a(s-\tau )}F(\tau ),
\label{eq133}
\end{equation}

with $r_{0}=r(t)|_{t=0}$ if we choose $t=0$.

\subsection{Bond price}

\subsubsection{Affine term structure}

The price at time $t$ of a zero-coupon bond $P(t,T,r)$ which pays one
currency unit at maturity $T$, $(0\leq t\leq T)$, is

\begin{equation}
P(t,T,r)=<\exp \{-\dint\limits_{t}^{T}dsr(s)\}>_{F},  \label{eq134}
\end{equation}

subject to terminal condition

\begin{equation}
P(T,T,r)=1,  \label{eq135}
\end{equation}

here $r(s)$ is given by Eq.(\ref{eq132}), $r=r(t)$ is the rate at time $t$,
and $<...>_{F}$ stands for the average with respect to all randomness
involved into the random force $F(t)$.

Further, substitution of $r(s)$ from Eq.(\ref{eq132}) into Eq.(\ref{eq134})
and straightforward integration over $ds$ yields

\begin{equation}
P(t,T,r)=\exp \left\{ A(t,T)-B(t,T)r(t)\right\} ,  \label{eq136}
\end{equation}

where the following notations have been introduced

\begin{equation}
B(t,T)=\dint\limits_{t}^{T}dse^{-a(s-t)}=\frac{1-e^{-a(T-t)}}{a},
\label{eq137}
\end{equation}

and

\begin{equation}
\exp \left\{ A(t,T)\right\} =<\exp
\{-\dint\limits_{t}^{T}dsB(s,T)F(s)\}>_{F}.  \label{eq138}
\end{equation}

Therefore, we conclude that the new model introduced by Eq.(\ref{eq131})
provides affine term structure.

To calculate the average in Eq.(\ref{eq138}), we use Eqs.(\ref{eq3}) and (%
\ref{eq13}) to obtain

\begin{equation}
\exp \left\{ A(t,T)\right\} =\exp \{\lambda
_{r}\dint\limits_{t}^{T}ds\dint\limits_{-\infty }^{\infty }d\eta p_{r}(\eta
)(e^{-\eta B(s,T)}-1)\},  \label{eq139}
\end{equation}

from which it follows directly that

\begin{equation}
A(t,T)=\lambda _{r}\dint\limits_{t}^{T}ds\dint\limits_{-\infty }^{\infty
}d\eta p_{r}(\eta )(e^{-\eta B(s,T)}-1).  \label{eq140}
\end{equation}

where $B(s,T)$ is defined by Eq.(\ref{eq137}), $\lambda _{r}$ is the number
of short term interest rate jumps per unit time, and $p_{r}(\eta )$ is the
probability density function of jump magnitudes.

We assume that probability density function $p_{r}(\eta )$ of short term
interest rate jump magnitudes is a normal distribution

\begin{equation}
p_{r}(\eta )=\frac{1}{\sqrt{2\pi }\delta _{r}}\exp \{-\frac{(\eta -\nu
_{r})^{2}}{2\delta _{r}^{2}}\},  \label{eq140p}
\end{equation}

where the market parameters $\nu _{r}$ and $\delta _{r}^{2}$ are the mean
and variance of short term interest rate jump magnitudes.

The calculation of the integral over $\eta $ in Eq.(\ref{eq140}) gives the
result

\begin{equation}
\dint\limits_{-\infty }^{\infty }d\eta p_{r}(\eta )(e^{-\eta B(s,T)}-1)=\exp
\{-\nu _{r}B(s,T)+\frac{\delta _{r}^{2}}{2}B^{2}(s,T)\}-1.  \label{eq141}
\end{equation}

Therefore, we have for $A(t,T)$

\begin{equation}
A(t,T)=\lambda _{r}\dint\limits_{t}^{T}ds\left\{ \exp \{-\nu _{r}B(s,T)+%
\frac{\delta _{r}^{2}}{2}B^{2}(s,T)\}-1\right\} .  \label{eq142}
\end{equation}

By introducing a new integration variable $y=B(s,T)$, we can express $A(t,T)$
in the form

\begin{equation}
A(t,T)=\lambda _{r}\dint\limits_{0}^{B(t,T)}\frac{dy}{1-ay}\exp \{-\nu _{r}y+%
\frac{\delta _{r}^{2}}{2}y^{2}-1\}.  \label{eq143}
\end{equation}

Thus, based on short term interest rate dynamics introduced by Eq.(\ref%
{eq131}), we obtained formula (\ref{eq136}) to price a zero coupon bond,
where $B(t,T)$ is given by Eq.(\ref{eq137}), and $A(t,T)$ is given by Eqs.(%
\ref{eq142}) or (\ref{eq143}).

\subsubsection{The term structure equation}

Defined by Eq.(\ref{eq134}) the price of zero-coupon bond $P(t,T,r)$ solves
the following integro-differential equation

\begin{equation}
\frac{\partial P(t,T,r)}{\partial t}-ar\frac{\partial P(t,T,r)}{\partial r}
\label{eq144}
\end{equation}

\begin{equation*}
+\lambda _{r}\dint\limits_{-\infty }^{\infty }d\eta p_{r}(\eta ){\LARGE \{}%
P(t,T,r+\eta )-P(t,T,r){\LARGE \}}=rP(t,T,r),
\end{equation*}

subject to terminal condition (\ref{eq135}), with $\lambda _{r}$ being the
number of short term interest rate jumps per unit time, $p_{r}(\eta )$ being
the probability density function of jump magnitudes, and parameter $a$ being
the speed at which the short term interest rate goes to its equilibrium
level at $t\rightarrow \infty $.

Following Vasi\v{c}ek \cite{Vasicek} terminology, we call Eq.(\ref{eq144})
"the term structure equation".

To find the solution to Eq.(\ref{eq144}) subject to terminal condition (\ref%
{eq135}) let's note that the affine term structure means that the bond price
admits solution of the form given by Eq.(\ref{eq136}). Substitution of Eq.(%
\ref{eq136}) into Eq.(\ref{eq144}) gives the two ordinary differential
equations to obtain $A(t,T)$

\begin{equation}
\frac{dA(t,T)}{dt}+\lambda _{r}\dint\limits_{-\infty }^{\infty }d\eta
p_{r}(\eta )\{e^{-\eta B(s,T)}-1\}=0,  \label{eq145}
\end{equation}

and $B(t,T)$

\begin{equation}
\frac{dB(t,T)}{dt}-aB(t,T)+1=0,  \label{eq146}
\end{equation}

subject to terminal conditions at $t=T$

\begin{equation}
A(T,T)=0,  \label{eq147}
\end{equation}

and

\begin{equation}
B(T,T)=0.  \label{eq148}
\end{equation}

To solve the system of equations (\ref{eq145}) and (\ref{eq146}) let's first
solve Eq.(\ref{eq146}). By separating the variables in Eq.(\ref{eq146}) we
write

\begin{equation*}
\frac{dB(t,T)}{aB(t,T)-1}=dt,
\end{equation*}

and

\begin{equation*}
\frac{1}{a}\ln |aB(t,T)-1|=t+C,
\end{equation*}

where $C$\ is an arbitrary constant of integration.

Hence, we have

\begin{equation*}
B(t,T)=\frac{1}{a}(1\pm e^{aC}e^{at}).
\end{equation*}

The value of constant $C$ and the sign can be fixed by imposing terminal
condition (\ref{eq148}), and finally, we come to the solution for $B(t,T)$
given by Eq.(\ref{eq137}).

Further, Eq.(\ref{eq145}) can be solved by straightforward integration.
Indeed, we have

\begin{equation*}
A(T,T)-A(t,T)=-\lambda _{r}\dint\limits_{t}^{T}ds\dint\limits_{-\infty
}^{\infty }d\eta p_{r}(\eta )\{e^{-\eta B(s,T)}-1\}.
\end{equation*}

It is obvious that after taking into account terminal condition (\ref{eq147}%
), we immediately come to the solution for $A(t,T)$ given by Eq.(\ref{eq140}%
).

Thus, we proved that the price of zero-coupon bond $P(t,T,r)$ defined by Eq.(%
\ref{eq134}) solves the new integro-differential term structure equation (%
\ref{eq144}).

\subsection{Vasi\v{c}ek model}

\subsubsection{Diffusion approximation}

To obtain the Vasi\v{c}ek bond pricing formula from Eqs.(\ref{eq136}), (\ref%
{eq137}) and (\ref{eq140}), we consider the diffusion approximation when $%
\nu _{r}\rightarrow 0$, $\delta _{r}^{2}\rightarrow 0$ and $\lambda
_{r}\rightarrow \infty $ while the products $\lambda
_{r}\dint\limits_{-\infty }^{\infty }d\eta p_{r}(\eta )\eta $ and $\lambda
_{r}\dint\limits_{-\infty }^{\infty }d\eta p_{r}(\eta )\eta ^{2}$ remain
finite.

It follows from Eq.(\ref{eq140}) that $A_{\mathrm{diff}}(t,T)$ in this
approximation is

\begin{equation}
A(t,T)\underset{\mathrm{diff}}{\rightarrow }A_{\mathrm{diff}}(t,T)
\label{eq149}
\end{equation}

\begin{equation*}
=\lambda _{r}\dint\limits_{-\infty }^{\infty }d\eta p_{r}(\eta )\{-\eta
B(s,T)+\frac{\eta ^{2}}{2}B^{2}(s,T)\}=-\lambda _{r}\nu _{r}B(s,T)+\frac{%
\sigma _{r}^{2}}{2}B^{2}(s,T),
\end{equation*}

where $B(s,T)$ is given by Eq.(\ref{eq137}), parameter $\nu _{r}$ is

\begin{equation}
\nu _{r}=\dint\limits_{-\infty }^{\infty }d\eta p_{r}(\eta )\eta ,
\label{eq150}
\end{equation}

and parameter $\sigma _{r}^{2}$ is

\begin{equation}
\sigma _{r}^{2}=\lambda _{r}\dint\limits_{-\infty }^{\infty }d\eta
p_{r}(\eta )\eta ^{2}.  \label{eq150s}
\end{equation}

\subsubsection{Long-term mean and instantaneous volatility}

Substituting $B(s,T)$ given by Eq.(\ref{eq137}) into Eq.(\ref{eq149}) and
evaluating the integral over $ds$ yield

\begin{equation}
A_{\mathrm{diff}}(t,T)=-\frac{\lambda _{r}\nu _{r}}{a}\left(
T-t-B(t,T)\right)  \label{eq151}
\end{equation}

\begin{equation*}
+\frac{\sigma _{r}^{2}}{2a^{2}}\left( T-t-2B(t,T)+\frac{1-e^{-2a(T-t)}}{2a}%
\right) .
\end{equation*}

Let us introduce the notation

\begin{equation}
b=\frac{\lambda _{r}\nu _{r}}{a}.  \label{eq152}
\end{equation}

Then we have

\begin{equation}
A_{\mathrm{diff}}(t,T)\equiv A_{\mathrm{Vasi\check{c}ek}}(t,T)=(b-\frac{%
\sigma _{r}^{2}}{2a^{2}})\left[ B(t,T)-(T-t)\right] -\frac{\sigma _{r}^{2}}{%
4a}B^{2}(t,T).  \label{eq153}
\end{equation}

A bond pricing equation (\ref{eq136}) in diffusion approximation has a form

\begin{equation}
P_{\mathrm{diff}}(t,T)=\exp \left\{ A_{\mathrm{Vasi\check{c}ek}%
}(t,T)-B(t,T)r(t)\right\} .  \label{eq154}
\end{equation}

We recognize in Eq.(\ref{eq154}) the Vasi\v{c}ek bond pricing formula \cite%
{Vasicek} with $A_{\mathrm{Vasi\check{c}ek}}(t,T)$ given by Eq.(\ref{eq153})
and $B(t,T)$ given by Eq.(\ref{eq137}). Hence, we conclude that $b$ defined
by Eq.(\ref{eq152}) is long term mean and $\sigma _{r}$ introduced by Eq.(%
\ref{eq150s}) is instantaneous volatility of the Vasi\v{c}ek short term
interest rate model. The parameter $\sigma _{r}^{2}/2a$ appearing \ in Eq.(%
\ref{eq153}) is sometimes called long term variance.

Let us remind that the Vasi\v{c}ek short term interest rate model has been
introduced by means of the following stochastic differential equation \cite%
{Vasicek}

\begin{equation}
dr=a(b-r)dt+\sigma _{r}dW(t),  \label{eq155}
\end{equation}

where $a$ is parameter $a$ ($a>0$) being the speed at which $r(t)$ goes to
its equilibrium level $b$ at $t\rightarrow \infty $, parameter $\sigma _{r}$
is instantaneous volatility and $W(t)$ is a standard Wiener process.

The parameters $a$, $b$, and $\sigma _{r}$ are positive constants. The
parameter $a$ in the Vasi\v{c}ek model has exactly the same meaning as in
our model introduced by Eq.(\ref{eq131}). The existence and meaning of the
phenomenological long term mean $b$ in the Vasi\v{c}ek model is explained in
our model by Eq. (\ref{eq152}) which emerges naturally in the diffusion
approximation. In other words, our model provides the quantitative
background to introduce parameter $b$ and lets us conclude that the long
term mean has stochastic dynamic origin. The origin of instantaneous
volatility $\sigma _{r}$ in the Vasi\v{c}ek model is explained in our model
by Eqs.(\ref{eq149}) and (\ref{eq150s}), that is, our model provides the
quantitative background to introduce the Vasi\v{c}ek \ parameter $\sigma
_{r} $.

Thus, we see that in diffusion approximation the short term interest rate
model defined by Eq.(\ref{eq131}) goes exactly into the well-known Vasi\v{c}%
ek model \cite{Vasicek}. The long term mean given by Eq.(\ref{eq152}) and
instantaneous volatility given by Eq.(\ref{eq150s}) are presented in terms
of $\lambda _{r}$ and the parameters involved into probability density
function $p_{r}(\eta )$ defined by Eq.(\ref{eq140p}). In other words, the
new stochastic model (\ref{eq131}) gives insight into stochastic dynamic
origin of the Vasi\v{c}ek long term mean $b$ and a short term interest rate
instantaneous volatility $\sigma _{r}$.

Let us emphasize that the stochastic dynamics (\ref{eq131}) initiates a
non-Gaussian probability distribution of short term interest rate in
contrast to the Vasi\v{c}ek model, which gives the Gaussian distribution.
However, despite the lack of Gaussianity, the stochastic model (\ref{eq131})
has the same degree of analytical tractability as the Vasi\v{c}ek model.

\subsubsection{Mean and variance}

It is interesting that the non-Gaussian model introduced by Eq.(\ref{eq131})
possesses exactly the same mean and variance as the Vasi\v{c}ek model, which
results in the Gaussian probability distribution of short term interest rate.

Indeed, it follows from Eq.(\ref{eq133}) that the mean $<r(s)>$ is

\begin{equation}
<r(s)>=e^{-as}r_{0}+\dint\limits_{0}^{s}d\tau e^{-a(s-\tau )}<F(\tau )>,
\label{eq156}
\end{equation}

and the variance is

\begin{equation}
\mathrm{Var(}r(s))=<(r(s)-<r(s)>)^{2}>  \label{eq157}
\end{equation}

\begin{equation*}
=<\left( \dint\limits_{0}^{s}d\tau e^{-a(s-\tau )}(F(\tau )-<F(\tau
)>)\right) ^{2}>,
\end{equation*}

where $<...>$ stands for the average defined by Eqs.(\ref{eq6})-(\ref{eq8}).

Taking into account Eq.(\ref{eq14}) and using Eq.(\ref{eq64}) we obtain for
the mean

\begin{equation}
<r(s)>=e^{-as}r_{0}+b(1-e^{-as})=b+(r_{0}-b)e^{-as},  \label{eq158}
\end{equation}

where $b$ is given by Eq.(\ref{eq152}).

To calculate the variance $\mathrm{Var(}r(s)$) let's use Eqs.(\ref{eq15})
and (\ref{eq64}) to find

\begin{equation}
\mathrm{Var(}r(s))=<(r(s)-<r(s)>)^{2}>=\frac{\sigma _{r}^{2}}{2a}%
(1-e^{-2as}),  \label{eq159}
\end{equation}

where $\sigma _{r}^{2}$\ is given by Eq.(\ref{eq150s}).

The conditional mean $<r(s)|r(t)>$ and variance $\mathrm{Var(}r(s)|r(t))$ are

\begin{equation}
<r(s)|r(t)>=b+(r(t)-b)e^{-a(s-t)},  \label{eq160}
\end{equation}

and

\begin{equation}
\mathrm{Var(}r(s)|r(t))=\frac{\sigma _{r}^{2}}{2a}(1-e^{-2a(s-t)}).
\label{eq161}
\end{equation}

The equations for $<r(s)|r(t)>$ and $\mathrm{Var(}r(s)|r(t))$ coincide
exactly with the Vasi\v{c}ek's equations (see, Eqs.(25) and (26) in \cite%
{Vasicek}) for conditional mean and variance of short term interest rate. It
has to be emphasized that we found an interesting coincidence between the
means and the variances of the non-Gaussian model introduced by Eq.(\ref%
{eq131}) and the Gaussian Vasi\v{c}ek model.

When $s\rightarrow \infty $, the conditional mean $<r(s)|r(t)>$ goes to the
long term mean $b$ given by Eq.(\ref{eq152}),

\begin{equation*}
\underset{s\rightarrow \infty }{\lim }<r(s)|r(t)>=b.
\end{equation*}

The conditional variance $\mathrm{Var(}r(s)|r(t))$ is increasing with
respect to $s$ from zero to the long term variance,

\begin{equation*}
\underset{s\rightarrow \infty }{\lim }\mathrm{Var(}r(s)|r(t))=\frac{\sigma
_{r}^{2}}{2a}.
\end{equation*}

Hence, the new stochastic model introduced by Eq.(\ref{eq131}) results in a
steady non-Normal probability distribution for $r(t)$ with the mean

\begin{equation*}
b=\frac{\lambda _{r}}{a}\dint\limits_{-\infty }^{\infty }d\eta p_{r}(\eta
)\eta ,
\end{equation*}

and the variance

\begin{equation*}
\frac{\sigma _{r}^{2}}{2a}=\frac{\lambda _{r}}{2a}\dint\limits_{-\infty
}^{\infty }d\eta p_{r}(\eta )\eta ^{2},
\end{equation*}

where $p_{r}(\eta )$ is defined by Eq.(\ref{eq140p}) and $\lambda _{r}$ is
the number of short term interest rate jumps per unit time.

\section{Generalized short term interest rate model}

\subsection{Vasi\v{c}ek model with shot noise}

Now we introduce a generalized short term interest rate model as a
superposition of the new model defined by Eq.(\ref{eq131}) and the Vasi\v{c}%
ek short term interest rate model given by Eq.(\ref{eq155})

\begin{equation}
dr=a(b-r)dt+\sigma _{r}dW(t)+F(t)dt,  \label{eq162}
\end{equation}

where all notations are the same as they were\ defined for Eqs.(\ref{eq131})
and (\ref{eq155}).

If we take into account that $dW(t)$ can be presented as $dW(t)=w(t)dt$,
where $w(t)$ is the white noise process, then Eq.(\ref{eq162}) reads

\begin{equation*}
dr=a(b-r)dt+\sigma _{r}w(t)dt+F(t)dt.
\end{equation*}

It is easy to see from Eq.(\ref{eq162}) that if at the time moment $t$ the
short term interest rate is $r(t)$ then later on, at the time moment $s$, $%
s>t$,\ it will be

\begin{equation}
r(s)=e^{-a(s-t)}r(t)+b(1-e^{-a(s-t)})  \label{eq163}
\end{equation}

\begin{equation*}
+\sigma _{r}\dint\limits_{t}^{s}d\tau e^{-a(s-\tau )}w(\tau
)+\dint\limits_{t}^{s}d\tau e^{-a(s-\tau )}F(\tau ),
\end{equation*}

or

\begin{equation}
r(s)=e^{-as}r_{0}+b(1-e^{-as})+\sigma _{r}\dint\limits_{t}^{s}d\tau
e^{-a(s-\tau )}w(\tau )+\dint\limits_{0}^{s}d\tau e^{-a(s-\tau )}F(\tau ),
\label{eq164}
\end{equation}

with $r_{0}=r(t)|_{t=0}$ if we choose $t=0$.

\subsection{Bond price: generalized formula}

The price at time $t$ of a zero-coupon bond $P(t,T,r)$ which pays one
currency unit at maturity $T$, $(0\leq t\leq T)$, is

\begin{equation}
P(t,T,r)=<\exp \{-\dint\limits_{t}^{T}dsr(s)\}>_{w,~F},\qquad \qquad
P(T,T,r)=1,  \label{eq165}
\end{equation}

where $r(s)$ is given by Eq.(\ref{eq163}) and $<...>_{w,~F}$ stands for the
average over white noise $w(t)$ and all randomness involved into shot noise $%
F(t)$.

Further, substitution of $r(s)$ from Eq.(\ref{eq163}) into Eq.(\ref{eq165})
and straightforward integration over $ds$ yields

\begin{equation}
P(t,T,r)=\exp \left\{ \mathcal{A}(t,T)-B(t,T)r(t)\right\} ,  \label{eq166}
\end{equation}

where $B(t,T)$ has been introduced by Eq.(\ref{eq137}), and

\begin{equation}
\exp \left\{ \mathcal{A}(t,T)\right\} =\exp \{-b(T-t-B(t,T))\}  \label{eq167}
\end{equation}

\begin{equation*}
\times <\exp \{-\sigma _{r}\dint\limits_{t}^{T}dsB(s,T)w(s)\}>_{w}<\exp
\{-\dint\limits_{t}^{T}dsB(s,T)F(s)\}>_{F},
\end{equation*}

because white noise and shot noise are independent of each other.

Therefore, we conclude that the generalized model introduced by Eq.(\ref%
{eq162}) provides affine term structure.

Further, we have (see, Eqs.(\ref{eq138}), (\ref{eq139}) and (\ref{eq142}))

\begin{equation}
<\exp \{-\dint\limits_{t}^{T}dsB(s,T)F(s)\}>_{F}  \label{eq168}
\end{equation}

\begin{equation*}
=\exp \left\{ \lambda _{r}\dint\limits_{t}^{T}ds\left\{ \exp \{-\nu
_{r}B(s,T)+\frac{\delta _{r}^{2}}{2}B^{2}(s,T)\}-1\right\} \right\} .
\end{equation*}

While for the average with respect to the white noise process \footnote{%
The characteristic functional $\Psi \lbrack \alpha (\tau )]$ of a white
noise process is
\par
\begin{equation}
\Psi \lbrack \alpha (\tau )]=<\exp \{i\dint\limits_{t}^{T}d\tau \alpha (\tau
)w(\tau )\}>_{w}=\exp \{-\frac{1}{2}\dint\limits_{t}^{T}d\tau \alpha
^{2}(\tau )\},
\end{equation}%
\par
where $\alpha (\tau )$ is an arbitrary sufficiently smooth function and $%
\left\langle ...\right\rangle _{w}$ stands for the average over white noise.
\par
To prove this equation we need to calculate $<\exp
\{i\dint\limits_{t}^{T}d\tau \alpha (\tau )w(\tau )\}>_{w}$. It can easily
be done if we write $\exp \{i\dint\limits_{t}^{T}d\tau \alpha (\tau )w(\tau
)\}$ as a power series expansion and take into account that white noise is a
random process which has two statistical moments only
\par
\begin{equation*}
\left\langle w(\tau )\right\rangle _{w}=0,\qquad \left\langle w(\tau
_{1})w(\tau _{2})\right\rangle _{w}=\delta (\tau _{1}-\tau _{2}).
\end{equation*}%
\par
{}} we have

\begin{equation}
<\exp \{-\sigma _{r}\dint\limits_{t}^{T}dsB(s,T)w(s)\}>_{w}=\exp \{\frac{%
\sigma _{r}^{2}}{2}\dint\limits_{t}^{T}dsB^{2}(s,T)\}  \label{eq170}
\end{equation}

\begin{equation*}
=\exp \left\{ \frac{\sigma _{r}^{2}}{2a^{2}}(T-t)-\frac{\sigma _{r}^{2}}{%
2a^{2}}B(t,T)-\frac{\sigma _{r}^{2}}{4a}B^{2}(t,T)\right\} .
\end{equation*}

It follows from Eqs.(\ref{eq167})-(\ref{eq170}) that

\begin{equation}
\mathcal{A}(t,T)=(b-\frac{\sigma _{r}^{2}}{2a^{2}})[B(t,T)-(T-t)]-\frac{%
\sigma _{r}^{2}}{4a}B^{2}(t,T)  \label{eq171}
\end{equation}

\begin{equation*}
+\lambda _{r}\dint\limits_{t}^{T}ds\dint\limits_{-\infty }^{\infty }d\eta
p_{r}(\eta )(e^{-\eta B(s,T)}-1),
\end{equation*}

which can be written as

\begin{equation}
\mathcal{A}(t,T)=A_{\mathrm{Vasi\check{c}ek}}(t,T)+A(t,T),  \label{eq172}
\end{equation}

with $A_{\mathrm{Vasi\check{c}ek}}(t,T)$ and $A(t,T)$ defined by Eqs.(\ref%
{eq153}) and (\ref{eq142}).

Thus, based on generalized short term interest rate dynamics introduced by
Eq.(\ref{eq162}), we obtained the formula (\ref{eq166}) for price of a zero
coupon bond, where $B(t,T)$ is given by Eq.(\ref{eq137}) and $\mathcal{A}%
(t,T)$ is given by Eq.(\ref{eq171}).

There are the following limit cases of Eq.(\ref{eq172})

\begin{equation}
\mathcal{A}(t,T)|_{\lambda =0}=A_{\mathrm{Vasi\check{c}ek}}(t,T),
\label{eq173}
\end{equation}

which results in the Vasi\v{c}ek formula (\ref{eq154}) for a bond price, and

\begin{equation}
\mathcal{A}(t,T)|_{b=0,~\sigma =0}=A(t,T),  \label{eq174}
\end{equation}

which results in bond pricing formula (\ref{eq136}) with $B(t,T)$ given by
Eq.(\ref{eq137}) and $A(t,T)$ given by Eqs.(\ref{eq142}).

Hence, the generalized short term interest rate dynamics (\ref{eq162})
describes the impact of interplay between Gaussian Geometric Brownian motion
and non-Gaussian Geometric Shot Noise on a bond price.

\subsubsection{Generalized term structure equation}

Defined by Eq.(\ref{eq165}) the price of zero-coupon bond $P(t,T,r)$ solves
the following integro-differential equation

\begin{equation}
\frac{\partial P(t,T,r)}{\partial t}+a(b-r)\frac{\partial P(t,T,r)}{\partial
r}+\frac{\sigma _{r}^{2}}{2}\frac{\partial ^{2}P(t,T,r)}{\partial r^{2}}
\label{eq175}
\end{equation}

\begin{equation*}
+\lambda _{r}\dint\limits_{-\infty }^{\infty }d\eta p_{r}(\eta ){\LARGE \{}%
P(t,T,r+\eta )-P(t,T,r){\LARGE \}}=rP(t,T,r),
\end{equation*}

subject to terminal condition (\ref{eq135}), with parameters $a$, $b$, and $%
\sigma _{r}$ being Vasi\v{c}ek's parameters in Eq.(\ref{eq155}), $\lambda
_{r}$ being the number of short term interest rate jumps per unit time, and $%
p_{r}(\eta )$ being the probability density function of jump magnitudes.

This is a generalized term structure equation in the case when the short
term interest rate solves the stochastic differential equation (\ref{eq162}).

To find the solution to Eq.(\ref{eq175}) subject to terminal condition (\ref%
{eq135}) let's note that the affine term structure means that the bond price
admits solution of the form given by Eq.(\ref{eq166}). Substitution of Eq.(%
\ref{eq166}) into Eq.(\ref{eq175}) gives the two ordinary differential
equations to obtain $\mathcal{A}(t,T)$ and $B(t,T)$. The ordinary
differential equation to find $\mathcal{A}(t,T)$ has a form

\begin{equation}
\frac{d\mathcal{A}(t,T)}{dt}+abB(t,T)+\frac{\sigma _{r}^{2}}{2}%
B^{2}(t,T)+\lambda _{r}\dint\limits_{-\infty }^{\infty }d\eta p_{r}(\eta
)\{e^{-\eta B(s,T)}-1\}=0,  \label{eq176}
\end{equation}

subject to terminal condition at $t=T$

\begin{equation}
\mathcal{A}(T,T)=0.  \label{eq177}
\end{equation}

For $B(t,T)$ we have Eq.(\ref{eq146}) subject to terminal condition given by
Eq.(\ref{eq148}), thus, we conclude that $B(t,T)$ is given by Eq.(\ref{eq137}%
). Then the solution to Eq.(\ref{eq176}) in terms of $B(t,T)$ is

\begin{equation}
\mathcal{A}(t,T)=\dint\limits_{t}^{T}ds\left\{ abB(t,T)+\frac{\sigma _{r}^{2}%
}{2}B^{2}(s,T)+\lambda _{r}\dint\limits_{-\infty }^{\infty }d\eta p_{r}(\eta
)(e^{-\eta B(s,T)}-1)\right\} ,  \label{eq178}
\end{equation}

which can be straightforwardly transformed into Eq.(\ref{eq171}).

Thus, it has been shown that the price of zero-coupon bond $P(t,T,r)$
defined by Eq.(\ref{eq165}) solves the generalized term structure equation (%
\ref{eq175}).

\section{Conclusion}

The unified framework consisting of two analytical approaches to value
options and bonds has been introduced and elaborated.

The options pricing approach is based on asset price dynamics that has been
modeled by the stochastic differential equation with involvement of shot
noise. It results in Geometric Shot Noise motion of asset price. A new
arbitrage-free integro-differential option pricing equation has been
developed and solved. New exact formulae to value European call and put
options have been obtained. The put-call parity has been proved. Based on
the solution of the option pricing equation, the new common Greeks have been
calculated. Three new Greeks associated with the market model parameters
have been introduced and evaluated. It has been shown that the developed
option pricing framework incorporates the well-known Black-Scholes equation 
\cite{Black-Scholes}. The Black-Scholes equation and its solutions emerge
from our integro-differential option pricing equation in the special case
which we call "diffusion approximation". The Geometric Shot Noise model in
diffusion approximation explains the stochastic dynamic origin of volatility
in the Black-Scholes model.

The generalized option pricing framework has been introduced and developed
based on asset price stochastic dynamics modeled by a superposition of
Geometric Brownian motion and Geometric Shot Noise. New formulae to value
European call and put options have been obtained as solutions to the
generalized arbitrage-free integro-differential pricing equation. Based on
these solutions new generalized Greeks have been introduced and calculated.

The bonds pricing analytical approach has been developed based on the
Langevin type stochastic differential equation with shot noise to model a
short term interest rate dynamics. It results in a non-Gaussian random
motion of short term interest rate. It is interesting that despite the lack
of normality, the new model possesses exactly the same mean and variance as
the Vasi\v{c}ek model, which results in the Normal distribution of short
term interest rate. A bond pricing formula has been obtained and it has been
shown that the model provides affine term structure. The new bond pricing
formula incorporates the well-known Vasi\v{c}ek solution \cite{Vasicek}. The
integro-differential term structure equation has been obtained and it has
been shown that the new bond pricing formula solves this equation. The
well-known Vasi\v{c}ek model for short term interest rate with its long-term
mean and instantaneous volatility comes out from our model in diffusion
approximation. The stochastic dynamic origin of the Vasi\v{c}ek long-term
mean and instantaneous volatility has been uncovered.

A generalized bond pricing model has been introduced and developed based on
short term interest rate stochastic dynamics modeled by a superposition of a
standard Wiener process and shot noise. A new bond pricing formula has been
found and it has been shown that the generalized model provides affine term
structure. The generalized integro-differential term structure equation has
been obtained and solved.

It has to be emphasized that despite the non-Gaussianity of probability
distributions involved, all newly elaborated quantitative models to value
options and bonds have the same degree of analytical tractability as the
Black--Scholes model \cite{Black-Scholes} and the Vasi\v{c}ek model \cite%
{Vasicek}. This circumstance allows one to derive new exact formulas to
value options and bonds.

The unified framework can be easily extended to cover valuation of other
types of options and a variety of financial products with option features
involved, and to elaborate enhanced short term interest rate models to value
interest rate contingent claims.

\section{Appendix A: Green's function method to solve the Black-Scholes
equation}

Let us show how to obtain the Black-Scholes formula to value a European call
option without converting the Black-Scholes equation into the heat equation.
In other words, we are aiming to get a straightforward solution to the
Black-Scholes problem given by Eqs.(\ref{eq92T}) and (\ref{eq93T}). Using
notation (\ref{eq17}), the Black-Scholes equation (\ref{eq92T}) with the
terminal condition given by Eq.(\ref{eq93T}) can be rewritten in a
mathematically equivalent form

\begin{equation}
-\frac{\partial C(x,T-t)}{\partial (T-t)}+\frac{\sigma ^{2}}{2}\frac{%
\partial ^{2}C(x,T-t)}{\partial x^{2}}+(r-q-\frac{\sigma ^{2}}{2})\frac{%
\partial C(x,T-t)}{\partial x}-rC(x,T-t)=0,  \label{eq179T}
\end{equation}

with the terminal condition given by Eq.(\ref{eq24T}).

The Green's function method is a convenient way to solve the problem given
by Eqs.(\ref{eq179T}) and (\ref{eq24T}). Green's function $%
G_{BS}(x-x^{\prime },T-t)$ of the Black-Scholes equation satisfies the
partial differential equation

\begin{equation}
\frac{\partial G_{BS}(x-x^{\prime },T-t)}{\partial (T-t)}=\frac{\sigma ^{2}}{%
2}\frac{\partial ^{2}G_{BS}(x-x^{\prime },T-t)}{\partial x^{2}}
\label{eq179}
\end{equation}

\begin{equation*}
+(r-q-\frac{\sigma ^{2}}{2})\frac{\partial G_{BS}(x-x^{\prime },T-t)}{%
\partial x}-rG_{BS}(x-x^{\prime },T-t)=0,
\end{equation*}

and the terminal condition

\begin{equation}
G_{BS}(x-x^{\prime },T-t)|_{t=T}=G_{BS}(x-x^{\prime },0)=\delta (x-x^{\prime
}).  \label{eq180}
\end{equation}

Having Green's' function $G_{BS}(x-x^{\prime },T-t)$, we can write the
solution of Eq.(\ref{eq179T}) with terminal condition Eq.(\ref{eq24T}) in
the form

\begin{equation}
C_{BS}(x,T-t)=K\dint\limits_{-\infty }^{\infty }dx^{\prime
}G_{BS}(x-x^{\prime },T-t)\max (e^{x^{\prime }}-1,0),  \label{eq181}
\end{equation}

where $C_{BS}(S,T-t)$ stands for the value of a European call option in the
Black-Scholes model.

Green's function introduced by Eq.(\ref{eq179}) with the terminal condition (%
\ref{eq180}) can be found by the Fourier transform method. With help of
definitions (\ref{eq27}) and (\ref{eq28}), equation (\ref{eq179}) reads

\begin{equation}
\frac{\partial G_{BS}(k,T-t)}{\partial (T-t)}=\left\{ -\frac{\sigma ^{2}k^{2}%
}{2}+ik(r-q-\frac{\sigma ^{2}}{2})-r\right\} G_{BS}(k,T-t),  \label{eq182}
\end{equation}

and the terminal condition (\ref{eq180}) for $G_{BS}(k,T-t)$ is

\begin{equation}
G_{BS}(k,T-t)|_{t=T}=G_{BS}(k,0)=1.  \label{eq183}
\end{equation}

The solution of the problem defined by Eqs.(\ref{eq182}) and (\ref{eq183}) is

\begin{equation}
G_{BS}(k,T-t)=e^{-r(T-t)}\exp \left\{ [-\frac{\sigma ^{2}k^{2}}{2}+ik(r-q-%
\frac{\sigma ^{2}}{2})](T-t)\right\} .  \label{eq184}
\end{equation}

Substitution of Eq.(\ref{eq184}) into Eq.(\ref{eq27}) gives us the Green's
function $G_{BS}(x-x^{\prime },T-t)$ of the Black-Scholes equation (\ref%
{eq179T})

\begin{equation*}
G_{BS}(x-x^{\prime },T-t)=\frac{e^{-r(T-t)}}{2\pi }\dint\limits_{-\infty
}^{\infty }dke^{ik(x-x^{\prime })}
\end{equation*}

\begin{equation}
\times \exp \left\{ [-\frac{\sigma ^{2}k^{2}}{2}+ik(r-q-\frac{\sigma ^{2}}{2}%
)](T-t)\right\}  \label{eq185}
\end{equation}

\begin{equation*}
=\frac{e^{-r(T-t)}}{\sigma \sqrt{2\pi (T-t)}}\exp \left\{ -\frac{%
[x-x^{\prime }+(r-q-\frac{\sigma ^{2}}{2})(T-t)]^{2}}{2\sigma ^{2}(T-t)}%
\right\} .
\end{equation*}

Then Eq.(\ref{eq181}) yields for the value of a European call option $%
C_{BS}(x,T-t)$

\begin{equation}
C_{BS}(x,T-t)=\frac{Ke^{-r(T-t)}}{\sigma \sqrt{2\pi (T-t)}}%
\dint\limits_{-\infty }^{\infty }dx^{\prime }\exp \left\{ -\frac{%
[x-x^{\prime }+(r-q-\frac{\sigma ^{2}}{2})(T-t)]^{2}}{2\sigma ^{2}(T-t)}%
\right\}  \label{eq186}
\end{equation}

\begin{equation*}
\times \max (e^{x^{\prime }}-1,0).
\end{equation*}

By introducing a new integration variable $z$, $x^{\prime }\rightarrow z$,

\begin{equation*}
z=\frac{x-x^{\prime }+(r-q-\frac{\sigma ^{2}}{2})(T-t)}{\sigma \sqrt{T-t}},
\end{equation*}

and using Eq.(\ref{eq90}) we find for a European call option $C_{BS}(S,T-t)$

\begin{equation}
C_{BS}(S,T-t)=\frac{Se^{-q(T-t)}}{\sqrt{2\pi }}e^{-\frac{\sigma ^{2}}{2}%
(T-t)}\dint\limits_{-\infty }^{d_{2}}dze^{-\frac{z^{2}}{2}}e^{-z\sigma \sqrt{%
T-t}}  \label{eq187}
\end{equation}

\begin{equation*}
-\frac{Ke^{-r(T-t)}}{\sqrt{2\pi }}\dint\limits_{-\infty }^{d_{2}}dz\exp
\left\{ -\frac{z^{2}}{2}\right\} ,
\end{equation*}

where parameter $d_{2}$ is given by

\begin{equation}
d_{2}=\frac{\ln \frac{S}{K}+(r-q-\frac{\sigma ^{2}}{2})(T-t)}{\sigma \sqrt{%
T-t}}.  \label{eq188}
\end{equation}

Introducing integration variable $y$, $y=z+\sigma \sqrt{T-t}$ in the first
term of Eq.(\ref{eq187}) yields

\begin{equation*}
C_{BS}(S,T-t)=\frac{Se^{-q(T-t)}}{\sqrt{2\pi }}\dint\limits_{-\infty
}^{d_{2}+\sigma \sqrt{T-t}}dye^{-\frac{y^{2}}{2}}
\end{equation*}

\begin{equation*}
-\frac{Ke^{-r(T-t)}}{\sqrt{2\pi }}\dint\limits_{-\infty }^{d_{2}}dz\exp
\left\{ -\frac{z^{2}}{2}\right\} .
\end{equation*}

It is easy to see from the equation above that we came to the well-known
solution to the Black-Scholes equation (\ref{eq92T})

\begin{equation*}
C_{BS}(S,T-t)=Se^{-q(T-t)}N(d_{1})-Ke^{-r(T-t)}N(d_{2}),
\end{equation*}

where $N(d)$ is the cumulative distribution function of the standard normal
distribution given by Eq.(\ref{eq100}) and parameter $d_{1}$ is

\begin{equation}
d_{1}=d_{2}+\sigma \sqrt{T-t}=\frac{\ln \frac{S}{K}+(r-q+\frac{\sigma ^{2}}{2%
})(T-t)}{\sigma \sqrt{T-t}},  \label{eq189}
\end{equation}

with $d_{2}$ defined by Eq.(\ref{eq188}).

\section{Appendix B: Greek $\Theta _{C}$ in diffusion approximation}

Here we present a walkthrough to calculate Greek $\Theta _{C}$ (see, Eq.(\ref%
{eq56})) in diffusion approximation.

It follows from Eqs.(\ref{eq37}) and (\ref{eq38}) that in diffusion
approximation

\begin{equation}
\lambda \varsigma \underset{\mathrm{diff}}{\rightarrow }\lambda \nu +\frac{%
\sigma ^{2}}{2},  \label{eqB1}
\end{equation}

and

\begin{equation}
\lambda \xi (k)\underset{\mathrm{diff}}{\rightarrow }i\lambda k\nu -\frac{%
k^{2}\sigma ^{2}}{2},  \label{eqB2}
\end{equation}

where parameter $\sigma $ has been introduced by Eq.(\ref{eq88}).

Using Eqs.(\ref{eqB1}), (\ref{eqB2}) and substituting $ike^{ikz}$ with $%
\partial e^{ikz}/\partial z$ we have

\begin{equation*}
\Theta _{C}\underset{\mathrm{diff}}{\rightarrow }\Theta _{C}^{\mathrm{diff}%
}=qSe^{-q(T-t)}N(d_{1})-rKe^{-r(T-t)}N(d_{2})+Se^{-q(T-t)}\lambda \varsigma
N(d_{1})
\end{equation*}

\begin{equation*}
-Se^{-q(T-t)}\frac{e^{-\lambda \varsigma (T-t)}}{\sigma \sqrt{2\pi (T-t)}}%
\dint\limits_{-\infty }^{l}dze^{-z}\left( \lambda \nu \frac{\partial }{%
\partial z}+\frac{\sigma ^{2}}{2}\frac{\partial ^{2}}{\partial z^{2}}\right)
\end{equation*}

\begin{equation}
\times \exp \left\{ -\frac{(z+\lambda \nu (T-t))^{2}}{2\sigma ^{2}(T-t)}%
\right\}  \label{eqB3}
\end{equation}

\begin{equation*}
+\frac{Ke^{-r(T-t)}}{\sigma \sqrt{2\pi (T-t)}}\dint\limits_{-\infty
}^{l}dz\left( \lambda \nu \frac{\partial }{\partial z}+\frac{\sigma ^{2}}{2}%
\frac{\partial ^{2}}{\partial z^{2}}\right)
\end{equation*}

\begin{equation*}
\times \exp \left\{ -\frac{(z+\lambda \nu (T-t))^{2}}{2\sigma ^{2}(T-t)}%
\right\} ,
\end{equation*}

where Eqs.(\ref{eq102}) and (\ref{eq105}) have been taken into account.

The integrals involved into Eq.(\ref{eqB3}) are

\begin{equation*}
\dint\limits_{-\infty }^{l}dze^{-z}\left( \lambda \nu \frac{\partial }{%
\partial z}+\frac{\sigma ^{2}}{2}\frac{\partial ^{2}}{\partial z^{2}}\right)
\exp \left\{ -\frac{(z+\lambda \nu (T-t))^{2}}{2\sigma ^{2}(T-t)}\right\}
\end{equation*}

\begin{equation}
=\lambda \nu e^{-l}\exp \left\{ -\frac{(l+\lambda \nu (T-t))^{2}}{2\sigma
^{2}(T-t)}\right\} +\lambda \nu \dint\limits_{-\infty }^{l}dze^{-z}\exp
\left\{ -\frac{(z+\lambda \nu (T-t))^{2}}{2\sigma ^{2}(T-t)}\right\}
\label{eqB4}
\end{equation}

\begin{equation*}
+\frac{\sigma ^{2}}{2}e^{-l}\frac{\partial }{\partial z}\left( \exp \left\{ -%
\frac{(z+\lambda \nu (T-t))^{2}}{2\sigma ^{2}(T-t)}\right\} \right) {\LARGE |%
}_{z=l}
\end{equation*}

\begin{equation*}
+\frac{\sigma ^{2}}{2}e^{-l}\exp \left\{ -\frac{(l+\lambda \nu (T-t))^{2}}{%
2\sigma ^{2}(T-t)}\right\}
\end{equation*}

\begin{equation*}
+\frac{\sigma ^{2}}{2}\dint\limits_{-\infty }^{l}dze^{-z}\exp \left\{ -\frac{%
(z+\lambda \nu (T-t))^{2}}{2\sigma ^{2}(T-t)}\right\} ,
\end{equation*}

and

\begin{equation*}
\dint\limits_{-\infty }^{l}dz\left( \lambda \nu \frac{\partial }{\partial z}+%
\frac{\sigma ^{2}}{2}\frac{\partial ^{2}}{\partial z^{2}}\right) \exp
\left\{ -\frac{(z+\lambda \nu (T-t))^{2}}{2\sigma ^{2}(T-t)}\right\}
\end{equation*}

\begin{equation}
=\lambda \nu \exp \left\{ -\frac{(l+\lambda \nu (T-t))^{2}}{2\sigma ^{2}(T-t)%
}\right\}  \label{eqB5}
\end{equation}

\begin{equation*}
=\frac{\sigma ^{2}}{2}\frac{\partial }{\partial z}\left( \exp \left\{ -\frac{%
(z+\lambda \nu (T-t))^{2}}{2\sigma ^{2}(T-t)}\right\} \right) {\LARGE |}%
_{z=l},
\end{equation*}

where parameter $l$ is given by Eq.(\ref{eq35}).

By substituting Eqs.(\ref{eqB4}) and (\ref{eqB5}) into Eq.(\ref{eqB3}) and
performing simple algebra we obtain

\begin{equation}
\Theta _{C}^{\mathrm{diff}}=qSe^{-q(T-t)}N(d_{1})-rKe^{-r(T-t)}N(d_{2})
\label{eqB6}
\end{equation}

\begin{equation*}
-Se^{-q(T-t)}\frac{e^{-\lambda \varsigma (T-t)}}{\sqrt{2\pi (T-t)}}\frac{%
\sigma }{2}e^{-l}\exp \left\{ -\frac{(l+\lambda \nu (T-t))^{2}}{2\sigma
^{2}(T-t)}\right\} .
\end{equation*}

Further simplification of Eq.(\ref{eqB6}) comes with the transformations of
the last term. Indeed, if we take into account that

\begin{equation*}
l+\lambda \nu (T-t)=x+(r-q-\frac{\sigma ^{2}}{2})(T-t),
\end{equation*}

and

\begin{equation*}
l+\lambda \varsigma (T-t)=x+(r-q)(T-t),
\end{equation*}

with $x$ given by Eq.(\ref{eq17}), then Eq.(\ref{eqB6}) reads

\begin{equation}
\Theta _{C}^{\mathrm{diff}}=qSe^{-q(T-t)}N(d_{1})-rKe^{-r(T-t)}N(d_{2})
\label{eqB7}
\end{equation}

\begin{equation*}
-\frac{\sigma Se^{-q(T-t)}}{2\sqrt{T-t}}N^{\prime }(d_{1}),
\end{equation*}

where $N^{\prime }(d)=\exp (-d^{2}/2)/\sqrt{2\pi }$ is derivative of $N(d)$
with respect to $d$, and $d_{1}$ is given by Eq.(\ref{eq189}).

Therefore, we came to the Black-Scholes Greek theta presented in Table 3.

\end{document}